\documentclass[pdflatex,sn-basic]{sn-jnl}

\usepackage{graphicx}%
\usepackage{multirow}%
\usepackage{amsmath,amssymb,amsfonts}%
\usepackage{amsthm}%
\usepackage{mathrsfs}%
\usepackage[title]{appendix}%
\usepackage{xcolor}%
\usepackage{textcomp}%
\usepackage{manyfoot}%
\usepackage{booktabs}%
\usepackage{listings}%

\usepackage[linesnumbered]{algorithm2e}
\usepackage{amsmath}
\usepackage{amssymb}

\newcommand{\dd}{\mathinner{.\,.}}

\newtheorem{theorem}{Theorem}
\newtheorem{lemma}[theorem]{Lemma}
\newtheorem{definition}{Definition}%

\newcommand{\no}[1]{}

\begin{document}

\title[Computing MEMs and Relatives on Repetitive Text
Collections]{\centering Computing MEMs and Relatives \\ on Repetitive Text Collections}

\author*[1,2]{\fnm{Gonzalo} \sur{Navarro}}\email{gnavarro@dcc.uchile.cl}
\affil[1]{Center for Biotechnology and Bioengineering (CeBiB), Chile}
\affil[2]{Department of Computer Science, University of Chile, Chile}

\abstract{
We consider the problem of computing the Maximal Exact Matches (MEMs) of a given
pattern $P[1\dd m]$ on a large repetitive text collection $T[1\dd n]$, which 
is represented as a (hopefully much smaller) run-length context-free grammar 
of size $g_{rl}$. We show that the problem can be solved in time 
$O(m^2 \log^\epsilon n)$, for any constant $\epsilon > 0$, on a data structure 
of size $O(g_{rl})$. Further, on a locally consistent grammar of size
$O(\delta\log\frac{n}{\delta})$, the time decreases to 
$O(m\log m(\log m + \log^\epsilon n))$. 
The value $\delta$ is a function of the substring
complexity of $T$ and $\Omega(\delta\log\frac{n}{\delta})$ is a tight lower
bound on the compressibility of repetitive texts $T$, so our structure has
optimal size in terms of $n$ and $\delta$.
We extend our results to several related problems, such as finding $k$-MEMs, 
MUMs, rare MEMs, and applications. 
}

\keywords{repetitive texts, substring complexity, grammar compression, 
locally consistent parsing, compressed data structures, MEMs, MUMs}

\maketitle

\section{Introduction and Related Work}

Inexact sequence matching is the norm in Bioinformatic applications. Mutations 
in the genomes, and even the possible errors that arise in the sequence 
acquisition 
process, makes researchers expect to see differences between the patterns they 
look for and what they call their occurrences \citep{Gus97,Ohl13,MBCT15}. For
example, when assembling a genome one must align the reads (which are
sequences obtained from the DNA of an individual, a few hundreds or thousands 
nucleotides long) to a reference genome (of length typically in the billions)
or a set thereof. For the reasons above, the read may
not appear in the genome in exact form, so one looks for the longest
substrings of the read that appear in the genome, to determine the most likely
positions where to align it. In pangenomics, one may compare the substrings of 
a whole gene or chromosome against another, or against a
set of genomes representative of a 
population, to find conserved regions and spot the places that differ, which 
may indicate genetic variations or diseases.

Those examples are applications of one of the most relevant tools for inexact
matching, which is finding the Maximal Exact Matches (MEMs) of a given pattern 
$P[1\dd m]$ in a text $T[1\dd n]$. A MEM is a maximal substring
$P[i\dd j]$ that appears in $T$ (i.e., $P[i-1\dd j]$ and $P[i\dd j+1]$ are 
out of bounds or do not occur in $T$). In the alignment of reads, $m$ is 
in the hundreds or thousands; when aligning a chromosome $m$ can surpass the 
millions. If the text is one genome, $n$ can be in the billions, but genomic
collections may contain hundreds to thousands of genomes.

In this paper we are interested in the case where $T$ is known in advance and 
thus can be indexed so that, later, we can efficiently find the MEMs of many 
patterns $P$ in it. This is the case in the two applications we have
mentioned; in the pangenomic case $T$ can be taken as the concatenation of all
the genomes in the collection. Many other applications of the problem of 
finding the MEMs of a new string in a static collection of strings are 
mentioned by \citet{Gus97}, \citet{Ohl13}, and \citet{MBCT15}, including 
simplified problems like finding longest common substrings
and equivalent problems like computing matching statistics.
some later.

Finding MEMs on an indexed text $T$ is a classic problem in stringology and 
can be solved in optimal $O(m)$ time using a suffix tree of $T$ \citep{Wei73,
McC76}. \citet[Sec.~7.8]{Gus97}, for example, shows how to solve the analogous
problem of computing matching statistics (we discuss this equivalence later). 
In applications where we handle massive texts $T$, however, suffix 
trees are too large to be maintained in main memory, even if they use linear 
space. The suffix tree of a single human genome, for example, whose length is
about 3 billion bases, may require 60GB of memory with a decent implementation.
This rules out suffix trees to represent genomes on the large bioinformatic 
collections that are arising, and make researchers look for alternatives using
less space. For example, \citet{OGK10} and \citet{BCKM13} use indices based on 
the Burrows-Wheeler Transform 
(BWT) \citep{BW94} to compute MEMs in time $O(m\log\sigma)$, where $\sigma$ is 
the alphabet size. Such indices take just a few GBs on a human genome, an
order of magnitude below the space required with suffix trees.

When considering large collections of genomes, however, even such sharp space 
reduction can be insufficient. Projects to sequence 100,000 human genomes have
been completed\footnote{\tt https://www.genomicsengland.co.uk/initiatives/100000-genomes-project}, and current projects aim to sequencing millions of human 
genomes\footnote{\tt https://b1mg-project.eu}. In that scenario, even the
BWT-based representations will need petabytes of main memory to run, or resort
to orders of magnitude slower disk storage.

A fortunate situation is that many of the fastest growing text collections,
including genome collections, are highly repetitive \citep{Navacmcs20.3}: two
genomes of the same species feature a small percentage of differences only.
Several text indices exploiting repetitiveness to reduce space have appeared 
\citep{Navacmcs20.2}. While probably not competitive with the BWT-based
indices we mentioned when indexing one genome, those indices may take orders of 
magnitude less space than the raw data, not just of its indices, when
representing a collection of many genomes. 

Those compressed indices support exact pattern matching, that is, they can 
list all the positions where $P$ occurs in $T$. While useful, this is 
insufficient to efficiently implement the MEM finding algorithms. This is the
problem we address in this paper.

\subsection{MEM finding with indices for repetitive text collections}

The classic MEM-finding algorithm runs on a suffix tree, and those that run on
BWT-based data structures emulate it. Compressed suffix trees for highly 
repetitive text collections do exist, but do not compress that much.
\citet{GNPjacm19} show how to simulate a suffix tree within space
$O(r\log\frac{n}{r})$, where $r$ is the number of equal-letter runs in the BWT 
of $T$. This representation can simulate the suffix-tree-based algorithm
in time $O(m\log\frac{n}{r})$ if we
run it backwards on $P$, using operations {\em parent} and {\em 
Weiner link} instead of {\em child} and {\em suffix link}. The problem is the
space: while $r$ is an accepted measure of repetitiveness \citep{KK20}, it is 
a weak one \citep{Navacmcs20.3,KK20}, and multiplying it by 
$\log\frac{n}{r}$ makes it grow by an order of magnitude.
Current implementations of compressed suffix trees for repetitive texts 
achieve remarkable space, but still use at least 2--4 bits per symbol 
\citep{RNO11,FGNPStcj18,CNic21,BCGHMNRalenex21}. 

Another trend has been to expand the functionality of a more basic compressed 
text index for repetitive texts so as to support specific operations, MEMs in 
our case. \citet{BGI20} show how to compute matching statistics 
(from where MEMs are easily extracted in $O(m)$ time) by extending the 
RLBWT-index \citep{MNSV09}, in $O(m(s+\log\log n))$ time and $O(r)$ space, with 
the help of a data structure that provides access to a symbol of $T$ in time 
$O(s)$. This can be, for example, the samples of the RLBWT-index, which add
$O(n/s)$ space to the index, or a context-free grammar of $T$, which provides
access in time $s=O(\log n)$ \citep{BLRSRW15}.
Various implementations of this idea \citep{ROLGB22,BGIKLMNPRdcc21.1,TFKG22}
showed its practicality on large genome collections, with indices that are
an order of magnitude smaller than the text.

All those results have been obtained on the so-called {\em suffix-based} 
compressed indices for repetitive collections \citep{Navacmcs20.2}. This is 
natural because those emulate variants of suffix
trees or arrays \citep{MM93}, which simplifies the problem of simulating the 
suffix tree traversal of the classic MEM-finding algorithm. Even the naive 
algorithm of searching for all the $O(m^2)$ substrings of $P$ can be run in
$O(m^2 \log\log n)$ time on those $O(r)$-sized indices. 

The problem is much harder on the so-called {\em parsing-based} indices 
\citep{Navacmcs20.2}. Those are potentially smaller than the suffix-based 
indices because they build on stronger measures of repetitiveness. For example,
the size $g$ of the smallest context-free grammar that generates $T$
is usually considerably smaller than $r$ \citep{Navacmcs20.3}. Because these
indices cut $T$ into phrases, even exact pattern matching is complicated
because the occurrences of $P$ can appear in many different forms, and many
possible cuts of $P$ must be tried out ($m-1$ in the general case)
\citep{CNPjcss21}. This makes the problem of finding MEMs considerably harder.
We are only aware of the results of \citet{Gao22}, who computes matching 
statistics in time $O(m^2 \log^\epsilon \gamma +m\log n)$ using 
$O(\delta\log\frac{n}{\delta})$ space (for any constant $\epsilon>0$), or 
$O(m^2+m\log\gamma
\log\log\gamma + m\log n)$ using $O(\delta\log\frac{n}{\delta} + \gamma\log
\gamma)$ space. Here $\delta \le \gamma$ are lower-bounding measures of 
repetitiveness \citep{KP18,CEKNPtalg20}. The size $O(\delta\log\frac{n}{\delta})$
matches a tight lower bound on the size of compressed representations of $T$ 
\citep{KNPtit23}, so a structure of this size uses
asymptotically optimal space for every $n$ and $\delta$.

\subsection{Our contribution}

We solve the MEM finding problem, and several relatives, within less space
and/or time than previous work, for the case of repetitive text collections.

Let $g_{rl}$ be the size of any run-length context-free grammar generating $T$ 
(those include and extend classic context-free grammars). The smallest such
grammar is of size $g_{rl} = O(\delta\log\frac{n}{\delta})$ \citep{KNPtit23}.
We first show that, on an index of size $O(g_{rl})$, one can compute the MEMs 
in time $O(m^2\log^\epsilon g_{rl})$, for any constant $\epsilon>0$. 
This is done by sliding the window $P[i\dd j]$ of the classic algorithm while 
we simulate the process of searching for that window with the grammar. The 
simulation is carefully crafted to avoid expensive operations, so the time
stays proportional to the number of cuts tried out on a single search for $P$. 
The space $O(g_{rl})$ is the least known to support direct access to 
$T$ with logarithmic time guarantees \citep{Navacmcs20.3}, so improving our
space is likely to involve breaking this long-standing barrier as well. 
Our result essentially matches the first one of \citet{Gao22}, which, although
he did not claim so, could also run within $O(g_{rl})$ space.

We further show that, on a particular grammar featuring local consistency 
properties \citep{KNOlatin22}, we can reduce the time to 
$O(m\log m(\log m + \log^\epsilon n))$ by exploiting the fact 
that only $O(\log(j-i+1))$ cuts need to be tried out for $P[i\dd j]$, and using
much more sophisticated techniques to amortize the costs. This grammar is of
size $O(\delta\log\frac{n}{\delta})$, which is optimal for every $n$ and 
$\delta$, and within this space we sharply break the quadratic time of previous
solutions that run in grammar-bounded space.

We then turn to consider several relatives of the MEM finding problem, 
adapting our main algorithm to solve them:

\begin{itemize}
\item 
A natural generalization of MEMs are $k$-MEMs, the maximal substrings 
of $P$ that appear at least $k$ times in $T$. Those identify 
the parts of $P$ that have sufficient support in $T$, for example regions of 
a gene that appear in most genomes of a collection, or regions in a reference
genome that have sufficient coverage in a set of reads. Even with $k$ given at 
query time, this problem is also easily solved in $O(m)$ time with a 
suffix tree \citep{Nav16}, but obtaining the same on grammars is not so direct.
We generalize our results on MEMs to find $k$-MEMs 
in time $O(km^2\log^\epsilon n)$ within $O(g_{rl})$
space, or in time $O(m\log m(\log m + k\log^\epsilon n))$ within 
$O(\delta\log\frac{n}{\delta})$ space. For $k = \omega(\log^2 n)$, we 
provide faster solutions that run in time $O(m^2 \log^{2+\epsilon} n)$ and
$O(g)$ space, or in time $O(m\log m\log^{2+\epsilon} n)$ and 
$O(\delta\log\frac{n}{\delta})$ space.
\item
A stricter version of MEMs are MUMs (maximal unique matches),
which are maximal matches that appear exactly once in $P$ and in $T$.
MUMs have various applications to sequence alignment
\citep{DKFPWS99,MBCT15,GRR22}.
Classical solutions using suffix trees \citep{Sun10} 
and suffix arrays \citep{Ohl13} compute MUMs in $O(m+n)$ time, though they
are easily seen to take $O(m)$ time if $P$ and $T$ are indexed
separately. MUMs are also computed in time $O(m\log\sigma)$ using a BWT-based
index \citep{BCKM13}. The only compressed-space solution for 
repetitive texts we know \citep{GRR22} computes MUMs in $O(r)$ space 
(plus a grammar on $T$) and $O(m\log n)$ time. We compute MUMs within the same 
space and time complexities as for finding the MEMs. 
\item 
A natural generalization of MUMs are rare MEMs \citep{OK08}, which also have
applications in whole genome alignment \citep[p.~419]{Ohl13}. We say
that a MEM is $k$-rare if it appears at most $k$ times in $P$ and in $T$, so
MUMs are $1$-rare MEMs. We show that $k$-rare MEMs can be computed within the
same space and time of $k$-MEMs.
\end{itemize}

We finally show, through applications to problems like data compression and
genome assembly, that our techniques open the door to using parsing-based
indices in stringology problems that had been addressed only through the more
powerful (but more space-consuming) suffix-based ones.  

A conference version of this paper \citep{Nav23} appeared in {\em Proc. CPM
2023}. In this journal version the key result, Section~\ref{sec:fasterMEM},
was largely rewritten due to simplifications, improvements, and filling
considerable gaps of the conference version. Several minor problems were 
fixed in other sections as well. A new set of extensions of the basic result 
to related problems, as well as applications of our result, are also included 
in Sections~\ref{sec:relatives} and \ref{sec:app}. Finally, the whole paper 
includes more detailed explanations, figures, and a better presentation.

The roadmap of the paper is as follows. In Section~\ref{sec:MEMs} we formally
define MEMs, the problem of finding them, and the variants we consider in the
paper. We also describe the algorithms for finding MEMs and how they are 
adapted to find their variants. In Section~\ref{sec:grammars} we describe
the basic grammar-based index for pattern matching, on which our simple 
solution builds. This simple solution, quadratic in $m$, is described in 
Section~\ref{sec:quadratic}. Section~\ref{sec:lcg} then describes the more
sophisticated locally consistent grammar we will use, and how the pattern
matching problem is solved on it. The more complex, subquadratic, algorithm to
find MEMs on locally consistent grammars is then described in
Section~\ref{sec:fasterMEM}. This is the central result of the article. 
Section~\ref{sec:relatives} shows how the tools we have developed can be used
to solve the related problems we have considered: $k$-MEMs, MUMs, $k$-rare
MEMs, and so on. We explore some direct and not so direct applications of our 
results in Section~\ref{sec:app}. We give our conclusions and future work 
directions in Section~\ref{sec:concl}.

\section{Maximal Exact Matches (MEMs) and Relatives} \label{sec:MEMs}

We use the classic notation on strings $S[1\dd n]$, so $S[i]$ is the $i$th
symbol of $S$, $S[i\dd j]$ denotes $S[i] \cdots S[j]$ (i.e., the concatenation
of symbols $S[i]$ to $S[j]$ and the empty string $\varepsilon$
if $i>j$), $S[\dd j] = S[1\dd j]$ and $S[i\dd] = S[i\dd n]$. The concatenation
of strings $S$ and $S'$ is denoted $S \cdot S'$.
We assume that the reader is 
familiar with the concepts related to suffix trees \citep{Wei73,McC76,CR02}. 

We start by defining the problems of finding MEMs and $k$-MEMs, and how to 
solve them on suffix trees.

\medskip

\begin{definition}
A {\em Maximal Exact Match (MEM)} of a pattern $P[1\dd m]$ in a string $T$ is a 
nonempty substring $P[i\dd j]$ that occurs in $T$, but in addition
\begin{itemize}
\item $i=1$ or $P[i-1\dd j]$ does not occur in $T$, and
\item $j=m$ or $P[i\dd j+1]$ does not occur in $T$.
\end{itemize}
\end{definition}

\medskip

\begin{definition}
A {\em $k$-MEM} of a pattern $P[1\dd m]$ in a string $T$ is a nonempty
substring $P[i\dd j]$ that occurs at least $k$ times in $T$, but in addition
\begin{itemize}
\item $i=1$ or $P[i-1\dd j]$ occurs fewer than $k$ times in $T$, and
\item $j=m$ or $P[i\dd j+1]$ occurs fewer than $k$ times in $T$.
\end{itemize}
\end{definition}

\medskip

\begin{definition}
Given a text $T[1\dd n]$ that can be preprocessed, the {\em MEM-finding
problem} is that of, given a pattern $P[1\dd m]$, return the range $(i,j)$ of 
each of its MEMs $P[i\dd j]$ in $T$, in increasing order of $i$ (and $j$). 
A position where each MEM occurs in $T$ must also be returned.
The {\em $k$-MEM finding problem} is defined analogously, for $k$ given at 
query time. 
\end{definition}

\medskip

\begin{algorithm}[t!]
	$i \gets 1$; $j \gets 0$ ; 

	$v \gets$ suffix tree root ; 

        \While{$j<m$}
           { \If{\rm $v$ has no child labeled $P[i]$}
		{ $i \gets i+1$ ; $j \gets j+1$ ; }
	     \Else
	        { \While{\rm $j < m$ and $v$ has a child labeled $P[j+1]$}
		     { $j \gets j+1$ ;
		       $v \gets \textrm{the child of }v\textrm{ by }P[j+1]$ ; 
		     }
		  \textbf{report} $(i,j)$ with some occurrence of $v$ ; 

		  \While{\rm $i \le j < m$ and $v$ has no child labeled $P[j+1]$}
		     { $i \gets i+1$ ;
		       $v \gets \textrm{the suffix link of }v$ ;
		     }
	        }
            }
\vspace{3mm}
\caption{Finding the MEMs of $P[1\dd m]$ in $T$ using the suffix tree of $T$.}
\label{alg:stmems}
\end{algorithm}

\begin{figure}[t]
\centering
\includegraphics[width=0.3\textwidth]{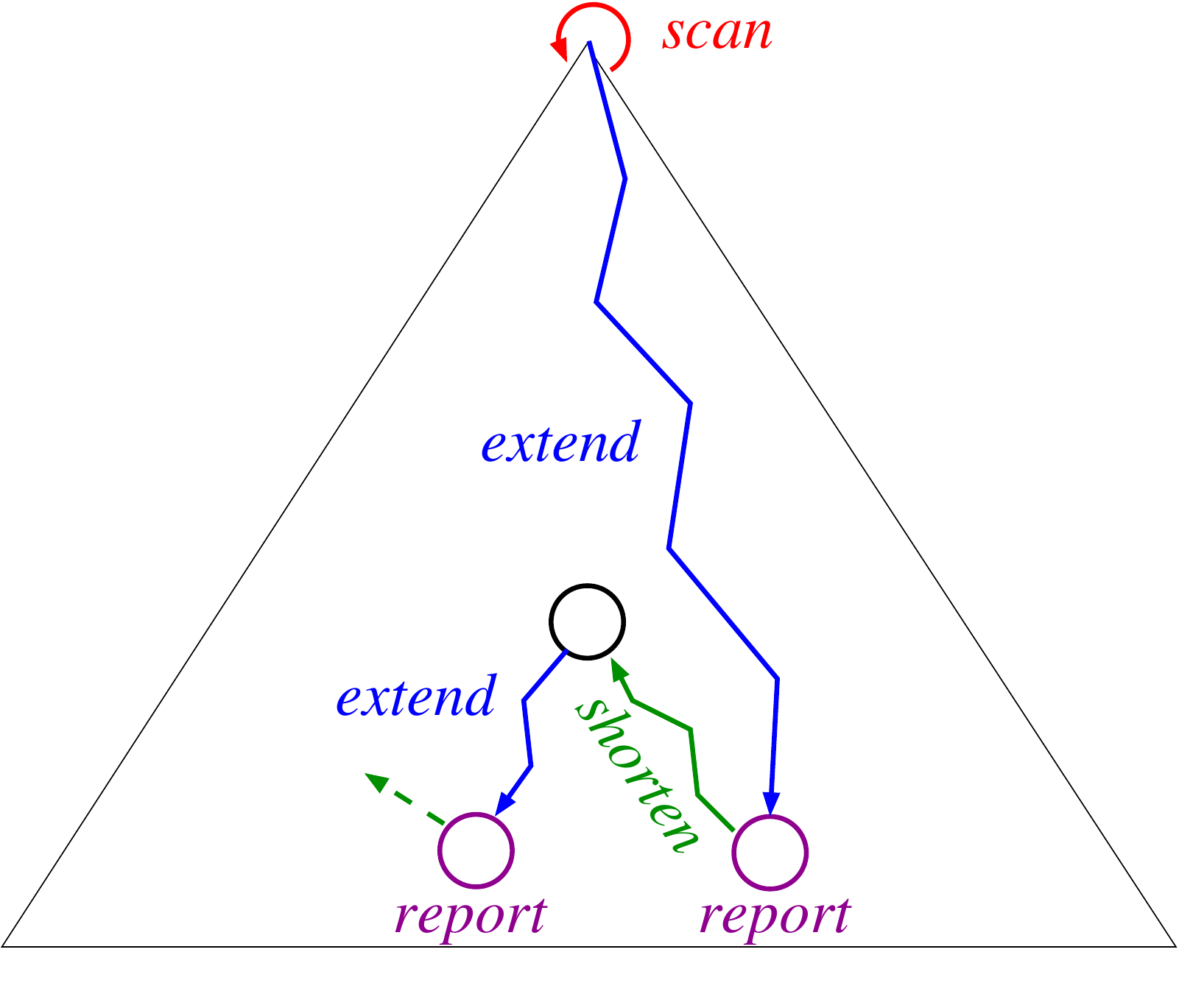}
\caption{Schematic view of the phases of the MEM finding algorithm on a suffix tree.}
\label{fig:stree}
\end{figure}

The MEM and $k$-MEM finding problems can be solved in $O(m)$ time with a 
suffix tree. Algorithm~\ref{alg:stmems} shows the solution for MEMs, 
abstracting away some complications of implementing it on the long edges of 
suffix trees. 

The algorithm slides a window $P[i\dd j]$ along $P$, maintaining the invariants
that (i) every MEM ending before $j$ has already been reported, (ii) the longest
suffix of $P[\dd j]$ that occurs in $T$ is $P[i\dd j]$, and (iii) $v$ is the 
suffix tree locus of $P[i \dd j]$. It iterates along four phases:
scanning, extending, reporting, and shortening; see Fig.~\ref{fig:stree}. The 
scanning phase (lines 4--6) considers the case $i=j+1$ (so $v$ is the suffix 
tree root), where $i$ and $j$ are increased as long as one cannot descend from 
$v$ by $P[j+1]$. The extending phase (lines 8--10) starts when we have found a 
child of the root by $P[i]$ and then descend from $v$ as much as possible by 
$P[i\dd j]$, increasing $j$. When we cannot further descend from the locus of 
$P[i\dd j]$ by $P[j+1]$, the invariants imply that $P[i\dd j]$ is a MEM, which 
is reported in line 11. Finally, the shortening phase (lines 12--14)
establishes the conditions that will allow increasing $j$ again, by increasing 
$i$ until $P[i\dd j+1]$ occurs in $T$ (or $i>j$). The tree suffix link is used 
to move from the locus of $P[i\dd j]$ to that of 
$P[i+1\dd j]$. Since each step increases $i$ or $j$, the total time is $O(m)$.

To adapt the algorithm to $k$-MEMs, we just ask that the suffix tree nodes we 
reach by $P[j+1]$ not only exist, but also that their subtrees contain $k$ 
leaves at least.

\medskip

Other notions related to MEMs are the MUMs and the $k$-rare MEMs.

\medskip

\begin{definition}
A {\em $k$-rare MEM} of a pattern $P[1\dd m]$ in a string $T$ is a nonempty
substring $P[i\dd j]$ that occurs in both $P$ and $T$, at most $k$ times 
in both cases. Further,
\begin{itemize}
\item $i=1$ or $P[i-1\dd j]$ does not occur in $T$, and
\item $j=m$ or $P[i\dd j+1]$ does not occur in $T$.
\end{itemize}
A $1$-rare MEM is called a {\em Maximal Unique Match (MUM)}.
\end{definition}

\medskip

The problem of finding the $k$-rare MEMs of $T$ is defined analogously as that
of finding the $k$-MEMs.
A variant of Algorithm~\ref{alg:stmems} to find the $k$-rare MEMs uses also a 
suffix tree of $P$, and traverses both suffix trees in synchronization, 
maintaining the locus of $P[i\dd j]$ in both suffix trees as it slides the
window $P[i\dd j]$. It reports $(i,j)$ in line 11 only if the number of leaves 
descending from both loci is at most $k$. The time is still $O(m)$.

\section{Grammar based Indices} \label{sec:grammars}

Let $T[1\dd n]$ be a text. Grammar-based 
compression of $T$ consists in replacing it by a context-free grammar (CFG) 
that generates only $T$ \citep{KY00}. The compression ratio is then the size of 
the grammar divided by the text size. 

We consider a slightly more powerful type of grammar called
run-length CFG (RLCFG), which includes run-length rules of 
constant size. We disallow rules of the form $A \rightarrow 
\varepsilon$, which are easily removed without increasing the grammar size.

\medskip

\begin{definition}
A {\em Run-Length Context-Free Grammar (RLCFG)} for $T$ is a context-free
grammar that generates (only) $T$, having exactly one rule per nonterminal
$A$. The rules are of the form $A \rightarrow B_1 \cdots B_t$ for $t>0$ and
terminals or nonterminals $B_i$ (this rule is said to be of size $t$), and
of the form $A \rightarrow B^t$ for $t>1$ and a terminal or nonterminal $B$, 
which is 
identical to $A \rightarrow B \cdots B$ with $t$ copies of $B$, but is said 
to be of size $2$. The {\em size} of the RLCFG is the sum of the sizes of 
its rules.
A {\em Context-Free Grammar (CFG)} for $T$ is a RLCFG for $T$ that does not 
use rules of the form $A \rightarrow B^t$.
\end{definition}

\medskip

Clearly, the size $g_{rl}$ of the smallest RLCFG for $T$ is always
less than or equal to the size $g$ of the smallest CFG for $T$.
Grammar-based compression (with or without run-length rules) has proved to be 
particularly effective on highly repetitive texts \citep{Navacmcs20.2}. While 
finding the smallest grammar is NP-hard \citep{CLLPPSS05}, heuristics like 
RePair obtain very good results \citep{LM00}.

\subsection{The grammar tree}

Since they have exactly one rule per nonterminal, the RLCFGs that generate
a single string $T$ have a unique parse tree, defined as follows
\citep[Sec.~4]{CEKNPtalg20}.

\medskip

\begin{definition}
The {\em parse tree} of a RLCFG for $T$ has a root labeled with the initial 
symbol. If a node is labeled $A$ and its rule is $A \rightarrow B_1\cdots B_t$,
then the node has $t$ children labeled $B_1,\ldots,B_t$ left to right. If its
rule is $A \rightarrow B^t$, then the node has $t$ children labeled $B$. 
The $i$th left-to-right leaf of the parse tree is thus labeled $T[i]$.
\end{definition}

\medskip

While the parse tree has size $\Omega(n)$, a convenient representation of a 
RLCFG is the so-called grammar tree, which is of size $O(g_{rl})$ 
\citep[Sec.~6]{CEKNPtalg20}.

\medskip

\begin{definition}
The {\em grammar tree} of a RLCFG is obtained by pruning its parse tree, 
preserving the leftmost internal node labeled $A$ for each nonterminal $A$, 
and converting the others to leaves. Further, for the remaining internal 
nodes labeled $A$ with rules $A \rightarrow B^t$ we preserve their first child
only, replacing the other $t-1$ children (which are leaves) with a single 
special leaf labeled $B^{[t-1]}$. If the RLCFG size is $g_{rl}$, its
grammar tree has $g_{rl}+1$ nodes.
\end{definition}

\medskip

\begin{figure}[t]
\centering
\includegraphics[width=0.7\textwidth]{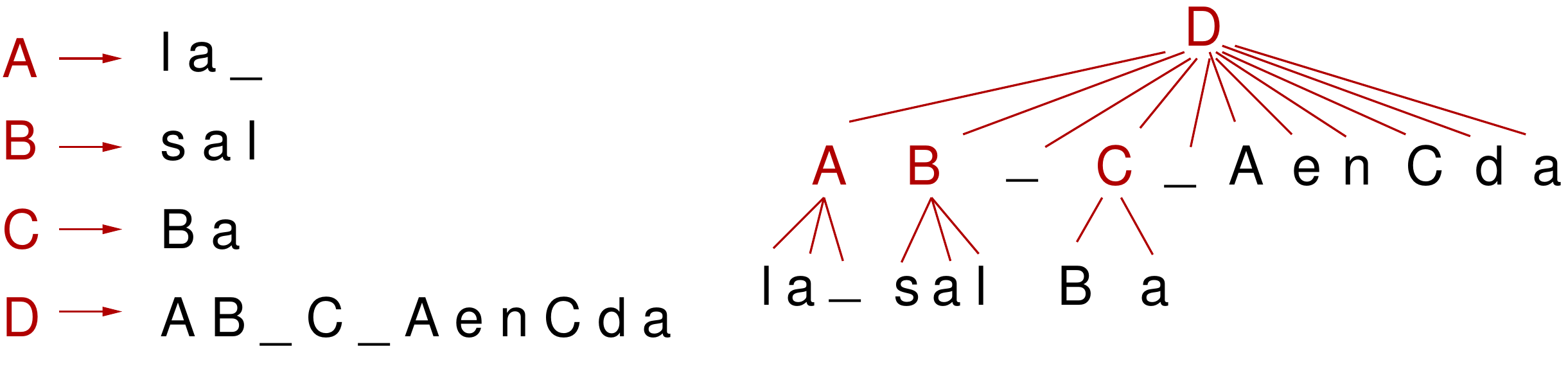}
\caption{A grammar for $T=\textsf{la\_sal\_sala\_la\_ensalada}$ and its grammar
tree. For example, $exp(\textsf{C})=\textsf{sala}$, and the parsing induced by
the grammar is $\textsf{l} \cdot \textsf{a} \cdot \textsf{\_} \cdot
\textsf{s} \cdot \textsf{a} \cdot \textsf{l} \cdot \textsf{\_} \cdot \textsf{sal} \cdot \textsf{a} \cdot
\textsf{\_} \cdot \textsf{la\_} \cdot \textsf{e} \cdot \textsf{n} \cdot \textsf{sala} \cdot
\textsf{da}$.}
\label{fig:grammartree}
\end{figure}

We will sometimes identify a nonterminal with its (only) internal node in the
grammar tree. Fig.~\ref{fig:grammartree} gives an example.

We call $exp(A)$ the string of terminals to which symbol $A$ expands, that is,
$exp(a)=a$ for terminals $a$, $exp(A)=exp(B_1)\cdots exp(B_t)$ if $A
\rightarrow B_1\cdots B_t$, and $exp(A)=exp(B)^t$ (i.e., $t$ concatenations of
$exp(B)$) if $A \rightarrow B^t$. The grammar tree defines a parse (or
partition into substrings called phrases) of $T$, as follows.

\medskip

\begin{definition}
The grammar tree, with leaves $v_1,\ldots,v_k$, induces the {\em parse}
$T = exp(v_1) \cdot exp(v_2) \cdots exp(v_k)$ into {\em phrases} $exp(v_i)$.
\end{definition}

\subsection{Primary occurrences}

A classic grammar-based index \citep{CNPjcss21} divides the occurrences of a 
pattern $P[1\dd m]$ into {\em primary} and {\em secondary}, depending on 
whether they cross a phrase boundary or lie within a phrase, respectively
(if $m=1$, its occurrences ending a phrase are taken as primary). The index
builds on the fact that every pattern $P$ has primary occurrences and that all
the secondary ones can be found inside pruned leaves of nonterminals that
contain other occurrences. The mechanism to find the primary occurrences is 
based on the parsing, but defined in a particular way to avoid reporting 
multiple times the primary occurrences that cross several 
phrase boundaries. The mechanism was extended to RLCFGs
\citep[Sec.~6 and App.~A]{CEKNPtalg20}.

\medskip

\begin{definition}
Let $\cal X$ and $\cal Y$ be multisets of strings defined as follows. For 
each rule $A \rightarrow B_1 \cdots B_t$, for each $1 < s \le t$, the string 
$exp(B_{s-1})^{rev}$ (i.e., $exp(B_{s-1})$ read backwards) is inserted in 
$\cal X$ and the string $exp(B_s)\cdots exp(B_t)$ is inserted in $\cal Y$; 
we say those two are {\em corresponding} strings. Similarly, for each rule
$A \rightarrow B^t$, $exp(B)^{rev}$ is inserted in $\cal X$ and
$exp(B)^{t-1}$ is inserted in $\cal Y$. A grid $\cal G$ has one
row per string in $\cal Y$ and one column per string in $\cal X$. After 
lexicographically sorting $\cal X$ and $\cal Y$, a point $(x,y)$ is set in
$\cal G$ if the $x$th string of $\cal X$ corresponds to the $y$th string of
$\cal Y$.
\end{definition}

\medskip

The grammar-based index includes a Patricia tree $P_{\cal X}$ storing the 
strings of $\cal X$ and another Patricia tree $P_{\cal Y}$ storing the strings 
of $\cal Y$ \citep{Mor68}. We add some data to the Patricia tree nodes for 
convenience. Each Patricia tree node $v$ stores the range 
$[v^1,v^2]$ of the left-to-right ranks of the leaves descending from $v$. 
The edges of the Patricia tree nodes can represent strings, so prefixes 
that end in the middle of an edge that leads to a node $v$ 
correspond to {\em virtual} nodes $u$; the range $[u^1,u^2]$ is the same 
$[v^1,v^2]$. The nodes $v$ also store their string depth $|v|$, which is also
easily computed for virtual nodes as we descend or ascend in the Patricia tree.
Finally, let $X=exp(B_{s-1})^{rev} \in \cal X$ be one string descending from 
node $x \in P_{\cal X}$, then we store $\langle x \rangle = B_{s-1}$ 
associated with $x$. Similarly,
a node $y \in P_{\cal Y}$ that prefixes $exp(B_s)\cdots exp(B_t) \in \cal Y$ 
stores $\langle y \rangle = B_s$ (from where we can obtain the subsequent 
siblings $B_{s+1}\cdots$). 

\begin{figure}[t]
\centering
\includegraphics[width=\textwidth]{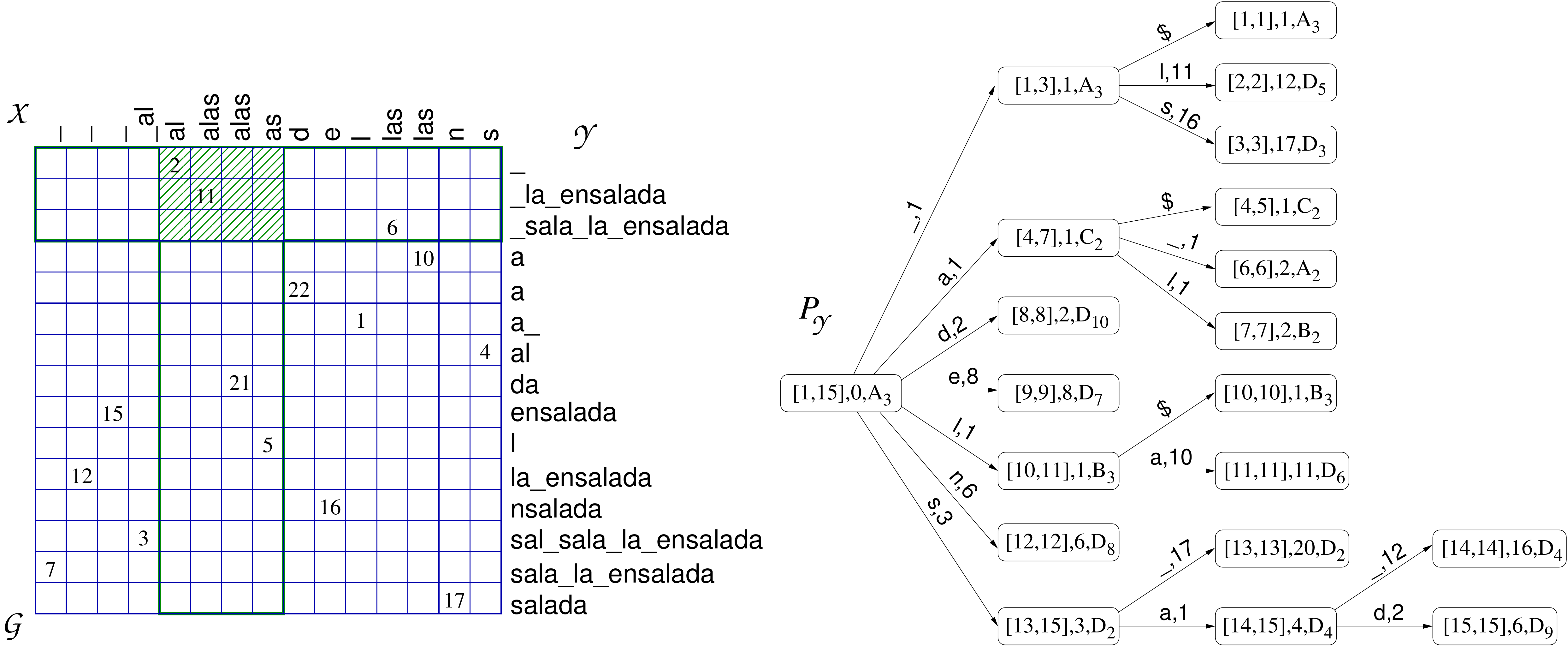}
\caption{The grid for the grammar of Fig.~\ref{fig:grammartree}, and the
ranges induced by the search for $P=\textsf{"a\_"}$. We obtain the position
$2$, which is inside the internal node for nonterminal \textsf{A} of the
grammar tree, and the position $11$, which is inside the internal node for
nonterminal \textsf{D} of the grammar tree (starting inside child \textsf{C}).
The occurrence at position $14$ is secondary. On the right, the Patricia tree
$P_{\cal Y}$. We write $[y^1,y^2],|y|,\langle y\rangle$ inside every node $y$.
We use $\$$ to denote the string terminator.}
\label{fig:grid}
\end{figure}

Each primary occurrence consists of a suffix of some string $X \in \cal X$ 
matching $P[1\dd r]$ corresponding to some string $Y \in \cal Y$ whose prefix 
matches $P[r+1\dd m]$, for some $1 \le r < m$ (if $m=1$, it is just a suffix
of $X$ matching $P$) \citep[Sec.~A.4]{CEKNPtalg20}. Therefore, to find the 
primary occurrences 
of $P$, the index tries out every cutting point $r$, and searches $P_{\cal X}$
for $P[1\dd r]^{rev}$ and $P_{\cal Y}$ for $P[r+1\dd m]$. If both nodes $x
\in P_{\cal X}$ and $y \in P_{\cal Y}$ exist, then the points in the orthogonal range 
$[x^1,x^2] \times [y^1,y^2]$ of $\cal G$ represent the primary occurrences of 
$P$ cut at position $r$, and are efficiently found with a geometric data 
structure on $\cal G$. The actual index stores a pointer to the node
$B_{s-1}$ in the grammar tree; we instead store the position $p$ of $T$
where $exp(B_{s-1})$ ends for such point, so we know that $P$ occurs in 
$T[p-r+1\dd p-r+m]$. See Fig.~\ref{fig:grid}.

Since $|{\cal X}|,|{\cal Y}| \le g_{rl}$, both the Patricia trees and the grid 
take $O(g_{rl})$ space. The index also needs to verify the matches of the 
Patricia trees. For this purpose it uses an $O(g_{rl})$-space
data structure $\cal A$ that can read, in $O(\ell)$ time, any length-$\ell$
prefix or suffix of $exp(A)$, for any nonterminal $A$
\citep[Lem~6.6]{CEKNPtalg20}. If $x$ is a node of $P_{\cal X}$, its 
corresponding string is the $|x|$-length reversed suffix of any string between 
the $x^1$th and the $x^2$th in $\cal X$. We can then obtain the string 
representing node $x$ by using $\cal A$ on $\langle x \rangle$; analogously
using $\cal A$ on $\langle y \rangle$ for nodes $y \in P_{\cal Y}$.

\subsection{Secondary occurrences} \label{sec:secondary}

Each primary occurrence found inside a nonterminal $A$ triggers a number of 
secondary occurrences:
\begin{enumerate}
\item Every other occurrence of $A$ in the grammar tree (which is necessarily
a leaf) contains an occurrence of $P$.
\item The occurrence in $A$ also occurs in the parent of $A$ in the grammar 
tree.
\end{enumerate}

From each primary occurrence in $A$, then, we recursively trigger searches to 
the next occurrence of $A$ and to the parent of $A$ in the grammar tree. If we
keep track of the offset of the occurrence within $exp(B)$ for each node $B$ in
the way, we can report the position of a secondary occurrence each time we reach
the root of the grammar tree. In the classic CFG-based index, this is shown
to amortize to constant time per secondary occurrence if each nonterminal is 
the root or appears at least twice in the grammar (the grammar can be modified
to enforce this condition) \citep{CNPjcss21}. The mechanism was extended to 
RLCFGs and to cases where the grammar cannot be modified 
\citep[Sec.~6 and A.4]{CEKNPtalg20}: each node points to its closest ancestor 
that is the root or occurs twice in the grammar tree, and the run-length 
nodes are treated in a special form to extract all the secondary occurrences
they encode.

A particularly relevant observation for this paper \citep{Nav18}
is that, on CFGs, the number of occurrences triggered by a primary occurrence 
depends only on that primary occurrence, and thus it can be associated with 
its point in the grid. 

\section{A Quadratic-Time Solution for MEMs} \label{sec:quadratic}

We now present a quadratic-time solution for finding MEMs that works with any 
RLCFG of size $g_{rl}$ for $T$. We use the $O(g_{rl})$-space data structures 
described in the previous section. Since every CFG is a particular case of 
RLCFG, our algorithm also runs on any CFG. 

The generic idea follows that of Algorithm~\ref{alg:stmems}, sliding a window 
$P[i\dd j]$ along the pattern. A trivial solution would be to use the classic
search described in Section~\ref{sec:grammars} for every window $P[i\dd j]$, 
to determine whether it occurs in $T$. Applied on any CFG of size $g$, such 
solution takes time $O((j-i+1)^2 + (j-i+1)\log^\epsilon g)$ for any constant 
$\epsilon>0$ if we resort to the classic index \citep[Cor.~1]{CNPjcss21}, 
because it needs to check every cutting 
point inside $P[i\dd j]$. This leads to $O(m^3 + m^2 \log^\epsilon g)$ time
complexity. 
Using the best current solution for arbitrary RLCFGs 
\citep[Thm.~A.4]{CEKNPtalg20}, this search takes $O((j-i+1)\log n)$ time,
leading to $O(m^2 \log n)$ total time for finding the MEMs of $P$ in $T$
within $O(g_{rl})$ space. In this section we will reduce this time only
slightly, but will set the grounds of a much better solution in the next
section.

The key aspect of the solution is to avoid redoing the search for every 
window. In particular, the trivial solution computes all the cuting points for 
every window $P[i\dd j]$, while two consecutive windows share most of them.
We instead maintain, along the process, a set of so-called {\em active 
positions} $r \in [i\dd j]$.

\medskip

\begin{definition} \label{def:active}
A position $r \in [i\dd j]$ is {\em active} on the window $P[i\dd j]$ iff
there is a primary occurrence of $P[r \dd j]$ in $T$ with the cut 
$P[r] \cdot P[r+1\dd j]$.
\end{definition}

\medskip

Note that, since we slide the window $P[i\dd j]$ forwards, once a position $r$
becomes inactive, it will not become active again. Note also that it is
possible that $r=j$ and thus $P[r+1\dd j]=\varepsilon$, which is needed to
find the MEMs of length $1$.

\subsection{Algorithm}

\begin{figure}[t]
\centering
\includegraphics[width=0.9\textwidth]{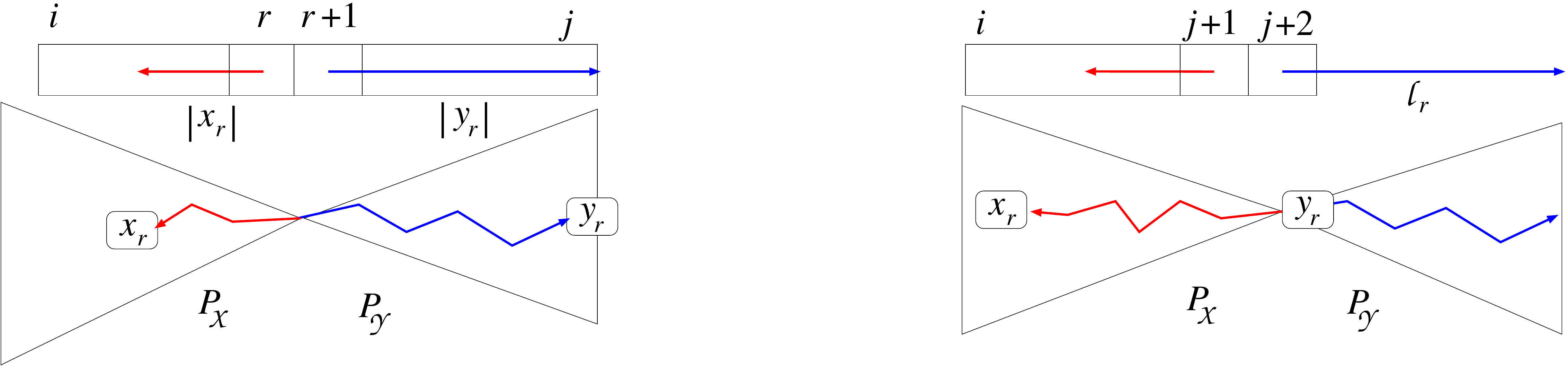}
\caption{On the left, the general relation between the nodes $x_r$ and $y_r$
for a general $r \in [i\dd j]$: while $|x_r|$ must be at most $r-i+1$ but it 
can be less, $|y_r|$ in $P_{\cal Y}$ must be exactly $j-r$; otherwise $r$ 
becomes inactive. 
On the right, the computation we carry out for a new $r = j+1$: we compute 
$\ell_r$, though $y_r$ corresponds to the root of $P_{\cal Y}$, and compute 
$x_r$. Later, as $j$ grows, $y_r$ will descend in $P_{\cal Y}$ and $x_r$ will 
possibly ascend in $P_{\cal X}$.}
\label{fig:R}
\end{figure}

The algorithm maintains invariants (i--ii) of Algorithm~\ref{alg:stmems}, 
that is, when the window is $P[i\dd j]$, every MEMs ending before $j$ has 
already been reported and the longest suffix of $P[\dd j]$ that occurs in $T$ 
is $P[i\dd j]$. Instead of the locus of $P[i\dd j]$ maintained in 
Algorithm~\ref{alg:stmems}, our algorithm maintains the set $R \subseteq 
[i\dd j]$ of active positions for $P[i\dd j]$, and for each such active 
position $r \in R$ it stores (see the left of Fig.~\ref{fig:R}):

\begin{itemize} 
\item The node $y_r \in P_{\cal Y}$ corresponding to $P[r+1\dd j]$; this node
can be virtual. Note that $[y_r^1,y_r^2]$ is the same range of 
rows in $\cal G$ of the strings of $\cal Y$ that start with $P[r+1\dd j]$.
\item The length $\ell_r$ of the maximum prefix of $P[r+1\dd]$ that prefixes
a string in $P_{\cal Y}$; note that $r$ is active iff $r+\ell_r \ge j$, so
$\ell_r$ can be zero if $r=j$.
\item The node $x_r \in P_{\cal X}$ corresponding to the longest prefix of 
$P[i\dd r]^{rev}$ that exists in $P_{\cal X}$, and such that there are points 
in $\cal G$ in the range $[x_r^1,x_r^2] \times [y_r^1,y_r^2]$. Note again that
$x_r$ can be virtual and that $[x_r^1,x_r^2]$ is the same range of columns in 
$\cal G$ of the strings of $\cal X$ that start with $P[r-|x_r|+1\dd r]^{rev}$.
Further, note we are interested only in values $r-|x_r|+1 \ge i$ because, by 
invariant (ii), $P[i-1\dd j]$ does not occur in $T$.
\end{itemize}

Our algorithm, depicted in Algorithm~\ref{alg:gmems}, iterates over $j$, from 
$0$ to $m-1$, and at each cycle it extends the current window so that it ends
in $j+1$. When $i=j+1$ (including when we start with $i=1$ and $j=0$), the 
window is empty and there are no active positions. This corresponds to the
scanning phase of Algorithm~\ref{alg:stmems}; the other phases are not as
clearly separated. Line 3 first sees if we can 
descend from the root of $P_{\cal X}$ by $P[j+1]$, so as to start a new 
nonempty substring $P[j+1,j+1]$. If this is not possible, it just increases
$i$ and goes for the next value of $j$. Otherwise, there will be active 
positions for the window ending at $j+1$ and we enter into the main process.

\begin{algorithm}[t]
	$i \gets 1$; $R \gets \emptyset$ ;  

        \For{$j \gets 0,\ldots,m-1$}
           { \lIf{$i=j+1$ \rm and the root of $P_{\cal X}$ has no child labeled $P[j+1]$}
		{ $i \gets i+1$ }
	     \Else
	        { $y_{j+1} \gets \textrm{root of } P_{\cal Y}$ ; 

		  $v \gets \textrm{descend in } P_{\cal Y}
			   \textrm{ as much as possible with } P[j+2\dd]$ ; 
		       $\ell_{j+1} \gets |v|$ ; 

		  $x_{j+1} \gets \textrm{descend in } P_{\cal X}
			   \textrm{ as much as possible with } P[i\dd j+1]^{rev}$ ; 

		  $r_{\min} \gets j+1$ ; 

		  \For{$r\in R$}
		      { \lIf{$r+\ell_r \le j$} { $R \gets R \setminus \{r\}$ }
		        \Else 
			   { $y_r \gets \textrm{child of } y_r
				  \textrm{ by } P[j+1]$ ; 

			     \lWhile{\rm range
					$[x_r^1,x_r^2] \times [y_r^1,y_r^2]$
				       is empty}
			         { $x_r \gets \textrm{(virtual) parent of } x_r$ }

			     \lIf{$|x_r|=0$} {$R \gets R \setminus \{r\}$ }
			     \lElseIf{$r-|x_r| < r_{\min}-|x_{r_{\min}}|$}
					{ $r_{\min} \gets r$ } 
			   }
		      }

		  $l \gets r_{\min}-|x_{r_{\min}}|+1$ ;

		  \If{$l>i$}
		     { \textbf{report} $(i,j)$ with position $T[p-j+i\dd p]$ ; 

		       $i \gets l$ 
		     }
		  \If{$i \le j+1$} 
		     { $p \gets j+1-r+\textrm{some text position in } 
				[x_{r_{\min}}^1,x_{r_{\min}}^2] \times
[y_{r_{\min}}^1,y_{r_{\min}}^2]$ ;

		       $R \gets R \cup \{ j+1 \}$ 
		     }
	        }
            }
	 \lIf{$i \le m$} {\textbf{report} $(i,m)$ with position $T[p-m+i\dd p]$ }
\vspace{5mm}
\caption{Finding the MEMs of $P[1\dd m]$ in $T$ using a grammar-based index.}
\label{alg:gmems}
\end{algorithm}

Lines 5--7
first create the new active position $r=j+1$, with the corresponding $y_r$ set 
at the root of $P_{\cal Y}$. To compute $\ell_r$, we descend in $P_{\cal Y}$
as much as possible by $P[r+1\dd]$. To compute $x_r$, we also descend in 
$P_{\cal X}$ as much as possible by $P[i \dd r]^{rev}$. Those are classic 
Patricia tree searches, first reaching a candidate node $v$ by comparing
only the branching characters in the trie, and then verifying which ancestor 
of $v$ is the correct answer. The verification proceeds by extracting the 
needed prefix from $\langle v \rangle$ in $P_{\cal Y}$ (at most $\ell_r+1$ 
characters) or the needed suffix in $P_{\cal X}$ (at most $|x_r|+1$ characters).
See the right of Fig.~\ref{fig:R}.

Note that, once we compute $\ell_r$, we know for every later $j$ that we can 
descend from $y_r$ by $P[j+1]$ iff $r+\ell_r \ge j+1$, and therefore we can 
compute the child node on the Patricia tree without accessing the text, both 
for explicit and virtual nodes $y_r$. Thus, by computing $\ell_r$ once when the
active position $r$ is created, in time $O(\ell_r)$, we save all the accesses 
to $T$ we would need to descend from virtual nodes $y_r \in 
P_{\cal Y}$. This would have been problematic because, when $y_r$ is not the 
root, its text position is not phrase-aligned, and thus we cannot access its
first symbols in constant time using $\cal A$.

Lines 8--17 then remove the positions that are no longer active and update
the variables of the surviving ones. Line 10 first removes the active
positions $r$ where $r+\ell_r = j$. On the remaining ones, each $y_r$ moves
to its child by $P[j+1]$ in $P_{\cal Y}$ in line 12, without accessing $T$
as explained. 

\begin{figure}[t]
\centering
\includegraphics[width=0.5\textwidth]{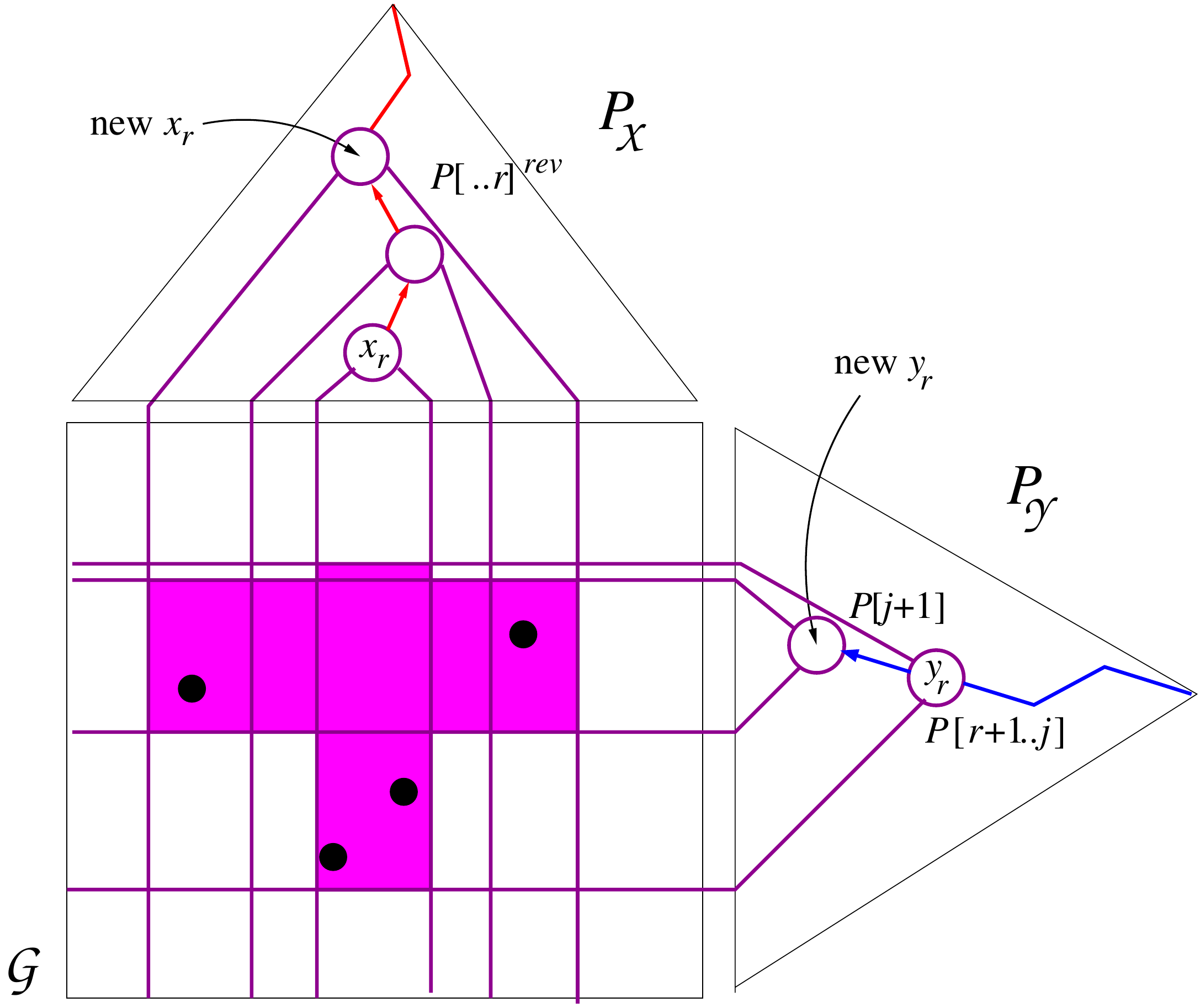}
\caption{Recomputing $x_r$ in $P_{\cal X}$ when $y_r$ descends in $P_{\cal
Y}$ by $P[j+1]$. If the range in $\cal G$ becomes empty, we move to the 
parent of $x_r$ (twice in this example) until it becomes nonempty again.}
\label{fig:retreat}
\end{figure}

Since moving down with $y_r$ may shrink the interval $[y_r^1,y_r^2]$,
line 13 updates the node $x_r$ accordingly, because the range 
$[x_r^1,x_r^2] \times [y_r^1,y_r^2]$ could now become empty.
For every active position $r$, as long as there are no points 
in $[x_r^1,x_r^2] \times [y_r^1,y_r^2]$, we move $x_r$ to its parent in 
$P_{\cal X}$. This process eventually terminates because, when $x_r$ is the 
root and $[x_r^1,x_r^2]$ is the whole range of columns, we know that there are 
points in the band $[y_r^1,y_r^2]$ because it corresponds to the node $y_r$.
Fig.~\ref{fig:retreat} illustrates the process.
If $x_r$ has become the root, however, $r$ is not active anymore per 
Definition~\ref{def:active}, so we remove it from $R$ in line 14.

Lines 8, 15, and 18 compute the active position $r_{\min}$ that yields 
occurrences starting at the leftmost position in the window, 
$l = r_{\min}-|x_{r_{\min}}|+1$. This minimization includes the 
active position $j+1$ we have not yet inserted in $R$.
It is then necessary to make $i$ grow to $l$ to re-establish invariant (ii) 
for $P[i\dd j+1]$. If $l>i$, then $P[i\dd j+1]$ does not occur in $T$
and thus $(i,j)$ was a MEM, by invariant (ii). Lines 20--21 then report the
MEM $(i,j)$ with its text position $p$ (collected in the previous cycle of $j$)
and increase $i$ to $l$, since only $P[l\dd j+1]$ occurs in $T$.
Increasing $i$ could make it exceed $j+1$ and thus make the window empty; 
otherwise line 24 collects the text position $p$ of some point in 
$[x_{r_{\min}}^1,x_{r_{\min}}^2] \times [y_{r_{\min}}^1,y_{r_{\min}}^2]$,
to be reported in case $(i,j+1)$ turns out later to be a MEM.
Finally, $j+1$ is inserted in $R$ as an active position.
Line 27 reports the final MEM when $j$ reaches
$m$, which again is correct by invariant (ii).

The geometric queries performed on $\cal G$ in line 13 are called (orthogonal) 
{\em range emptiness queries}. These tell whether there is any point in a given
rectangle of the grid. The other geometric queries are those of line 24, where 
we ask, in case a rectangle is not empty, for some point within it (and we 
extract the text position $p$ from the satellite data of the point). This 
second query is generally a subproduct of the first, so both can be seen as 
the same query.

\subsection{Analysis} 

For each value of $j$, Algorithm~\ref{alg:gmems} spends $O(1)$ time per active 
position. Since there are $O(m)$ active positions at any time, this amounts to 
$O(m^2)$ time. 

The exceptions are lines 6, 7, and 13, which are better charged to each active 
position $r$, from its creation to its inactivation. When $r$ is created, we 
spend $O(m)$ time to compute $\ell_r \le m$ and $x_r$ (since $|x_r| \le m$). Later, we can decrease 
$|x_r|$ several times, performing one range emptiness query in 
$[x_r^1,x_r^2] \times [y_r^1,y_r^2]$ per decrement of $|x_r|$ (in fact we 
can go directly to the lowest physical ancestor of $x_r$ rather than to its 
possibly virtual parent node, as otherwise the range $[x_r^1,x_r^2]$ will 
not change). Thus, we perform overall $O(m^2)$ emptiness queries, up to 
$m$ per position $r$ along its life. Maintaining the variables associated with
active positions allows us amortizing these costs along the process.

Emptiness queries on $\cal G$ can be solved in $O(\log^\epsilon g_{rl})$ time 
and $O(g_{rl})$ space for any constant $\epsilon>0$ \citep{CLP11}. 
The same complexity holds for returning one point in nonempty ranges. The 
total cost of the MEM finding algorithm is then $O(m^2\log^\epsilon g_{rl})$,
dominated by the $O(m^2)$ geometric queries. 

The grammar data structures we use are those of the classic solution
\citep{CNPjcss21} (we do not obtain better times in this section by using the
more sophisticated index of \citet{CEKNPtalg20}). Most of the construction 
cost is dominated by that of sorting ${\cal X}$ and ${\cal Y}$ in order to 
build the Patricia trees. Those sets can be sorted in time $O(g_{rl}\log^2 n)$ 
time and $O(g_{rl}\log n)$ space \citep[Sec.~3 \& 5.1]{CNPjcss21}, or in $O(n)$
time and space using the suffix arrays of $T$ and $T^{rev}$ 
\citep{MM93,KSB06}. While those construction times have been given for 
CFGs, they also apply to RLCFGs because we only introduce two strings for the 
rules of the form $A \rightarrow B^t$, and use general sorting algorithms.
The other main component of the construction cost is that of the data 
structures that support the geometric queries on ${\cal G}$, which can be 
built in time $O(g_{rl}\sqrt{\log g_{rl}})$ \citep{BP16}.

\medskip

\begin{theorem} \label{thm:mem-m2}
Assume we have a RLCFG of size $g_{rl}$ that generates only $T[1\dd n]$.
Then, for any constant $\epsilon>0$, we can build a data structure of size 
$O(g_{rl})$ that finds the MEMs of any given pattern $P[1\dd m]$ in time 
$O(m^2\log^\epsilon g_{rl}) \subseteq O(m^2\log^\epsilon n)$,
returning an occurrence of each MEM. The data structure can be built in 
$O(g_{rl}\log^2 n)$ time and $O(g_{rl}\log n)$ space, or in 
$O(n+g_{rl}\sqrt{\log g_{rl}})$ time and $O(n)$ space.
The query process uses $O(m)$ additional space on top of the data structure.
\end{theorem}

\medskip

If we use the smallest RLCFG that generates $T$, then it holds that 
$g_{rl} = O(\delta\log\frac{n}{\delta})$ \citep{Navacmcs20.3}, which matches
a lower bound on the space needed to represent repetitive strings (see 
next section for details). With this bound, the time of the algorithm can be 
written as $O(m^2(\log^\epsilon \delta + \log\log n))$. On the other hand, by 
using a range emptiness data structure of size $O(g_{rl}\log\log g_{rl})$ 
\citep{CLP11}, then the index is slightly larger but we can find the MEMs in 
time $O(m^2\log\log g_{rl})$.

\section{Indexing Locally Consistent Grammars (LCGs)}
\label{sec:lcg}

Before entering into the details of our more sophisticated solution for 
finding MEMs, we must
introduce some new concepts. A {\em locally consistent grammar (LCG)} is a kind of
RLCFG that guarantees that equal substrings of $T$ are covered by similar 
subtrees of the parse tree, differing in $O(1)$ nonterminals at each level
of both subtrees. LCGs have been used to produce grammar-based indices that
find all the primary occurrences with only a logarithmic number of cuts in $P$,
thereby obtaining exact pattern searches in time that grows only linearly with 
$m$ \citep{CEKNPtalg20,KNOlatin22,KNOarxiv23}. In this paper we make use of the 
latest result \citep{KNOlatin22,KNOarxiv23}. We
present a lighter informal description; see the original paper for full details.

Since LCGs are not necessarily the smallest RLCFGs that generate a given text
$T$, we will use another goal for the space achieved by a LCG \citep{RRRS13,CEKNPtalg20}.

\medskip

\begin{definition}
Given a text $T[1\dd n]$ we define, for every $\ell > 0$, $T_\ell$ as the 
number of distinct length-$\ell$ substrings in $T$, and then
$$\delta ~=~ \delta(T) ~=~ \max \{ T_\ell/\ell, ~\ell > 0 \}.$$ 
\end{definition}

The set of values $T_\ell$ is called the {\em substring complexity} of $T$,
and $\Omega(\delta\log\frac{n\log|\Sigma|}{\delta\log n}) \subseteq
\Omega(\delta\log\frac{n}{\delta})$, where $\Sigma$ is the alphabet of $T$,
is a lower bound to the space (measured in $\log(n)$-bit words) needed to
distinguish $T$ from the family of all strings with the same
$n$, $|\Sigma|$, and $\delta$ \citep{KNPtit23}. We then aim at using space
$O(\delta\log\frac{n\log|\Sigma|}{\delta\log n})$ for a grammar (and for an 
index), as this is asymptotically optimal. We note that, for every $T$, there
is a RLCFG of size $g_{rl} = O(\delta\log\frac{n\log|\Sigma|}{\delta\log n})$
generating $T$ \citep{KNOarxiv23}, so the size of the smallest RLCFG is a
finer lower bound than this $\delta$-based measure. Therefore, the index of 
Theorem~\ref{thm:mem-m2} is potentially smaller, yet slower, than the one we will
develop in this section. We note that, although the space we will achieve is
$O(\delta\log\frac{n\log|\Sigma|}{\delta\log n})$, we will use for simplicity
the coarser bound $O(\delta\log\frac{n}{\delta})$.

\subsection{The grammar}

We first define the grammar \citep[Sec.~3]{KNOlatin22}, which is produced
level by level, for $O(\log n)$ levels. Let $\Sigma$ be the alphabet of the
text $T[1\dd n]$ and also the set of terminals in the grammar. Let $S_k$ be 
the sequence of terminals and nonterminals forming level $k$ of the grammar.
Let $\ell_k = (4/3)^{\lceil k/2\rceil-1}$, and let $\mathcal{A}_k$ be
the set of symbols $A$ such that $|exp(A)| \le \ell_k$. Those are the 
symbols that can be grouped to form new nonterminals in level $k$; the
others are said to be ``paused'' in that level.

\begin{figure}[t]
\centering
\includegraphics[width=0.7\textwidth]{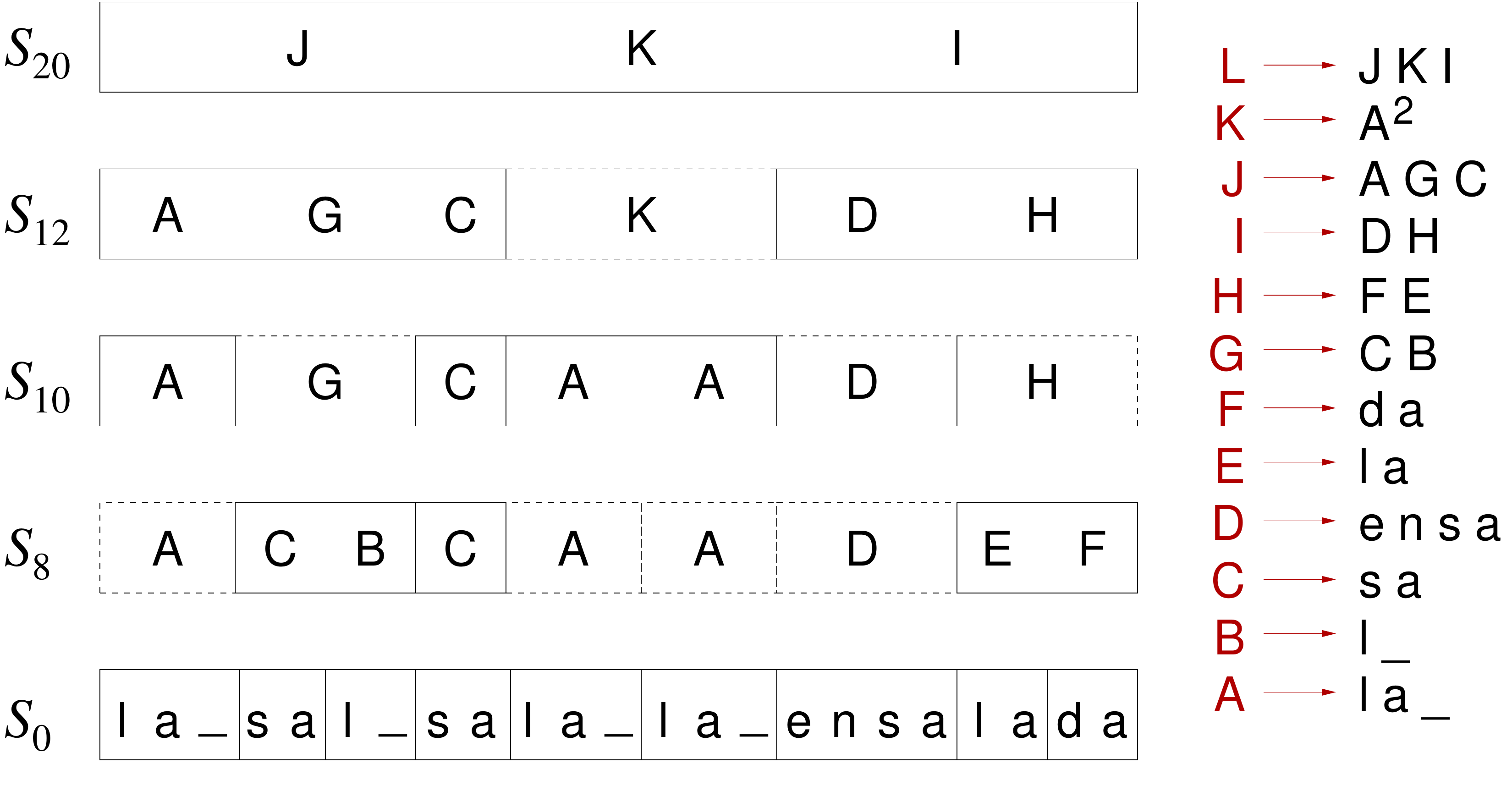}
\caption{Our LCG built over $T=\textsf{"la\_sal\_sala\_la\_ensalada"}$. We use
the lexicographic values of terminals and nonterminals as their permutation,
and show with rectangles the formed blocks (the underscore is taken as the
smallest terminal).
We show the levels $S_k$ where the value of $\ell_k$ allows changes in the 
parsing; symbols that cannot yet be grouped are in dashed rectangles.}
\label{fig:lcg}
\end{figure}

Our string at level $0$ is $S_0 = T$. To form the string $S_1$,
we detect the maximal {\em runs} of (at least 2) equal consecutive symbols in 
$S_0$ that are in $\mathcal{A}_1 = \Sigma$. For each such run, say of $t$ 
symbols $a\in\mathcal{A}_1$, we create the rule $A \rightarrow a^t$ and replace
the run by the nonterminal $A$. The resulting sequence after all the runs have 
been replaced is $S_1=rle_1(S_0)$, which contains terminals and 
nonterminals. To form level $2$, we 
define a function $\pi_2$ that reorders at random the distinct symbols of $S_1$,
and use it to define {\em blocks} in $S_1$. Each ``local minimum'' 
position $0<i<|S_1|$ such that 
$$\pi_2(S_1[i-1]) > \pi_2(S_1[i]) < \pi_2(S_1[i+1])$$
terminates a block. We also set ends of blocks at $|S_1|$ and before and 
after every symbol not in $\mathcal{A}_2$ (which is still $\Sigma$ per the
formula of $\ell_k$, so the runs introduced in $S_1$ cannot yet be grouped). 
For each distinct resulting block 
$S_1[i\dd j]$ we create a new rule $A \rightarrow S_1[i\dd j]$ and replace 
every occurrence of the same block in $S_1$ by $A$. The resulting string is 
called $S_2 = bc_2(S_1)$. The process continues in the 
same way for odd and even levels:
\begin{eqnarray*}
S_k  = & rle_k(S_{k-1}) & \textrm{if } k \textrm{ is odd}, \\
S_k  = & bc_k(S_{k-1}) & \textrm{if } k \textrm{ is even},
\end{eqnarray*}
until we reach $|S_k|=1$ for some $k=O(\log n)$; see Fig.~\ref{fig:lcg}. 
The algorithm is Las Vegas
type, trying out functions $\pi_k$ until obtaining some desired grammar size;
otherwise any functions $\pi_k$ yield a correct grammar of size $O(n)$ and
height $O(\log n)$.
\citet{KNOlatin22} prove that, after $O(1)$ attempts (i.e., in $O(n)$ expected 
time), a RLCFG of asymptotically optimal size 
$O(\delta\log\frac{n}{\delta})$ is obtained.

A key property of this grammar is {\em local consistency}.
Let $\mathcal{B}_k$ be the set of all the ends of level-$k$ blocks:
$$\mathcal{B}_k ~=~ \{ |exp(S_k[\dd j])|,~ 1 \le j \le |S_k| \},$$
where we are extending $exp(\cdot)$ homomorphically to strings.
The cuts of level $k$ that fall inside the substring at $T[i\dd j]$ have the 
following positions inside $T[i\dd j]$:
$$\mathcal{B}_k(i,j) ~=~ \{ p-i+1,~p \in \mathcal{B}_k \cap [i\dd j-1] \}.$$
Local consistency makes the sets $\mathcal{B}_k(i,j)$ and 
$\mathcal{B}_k(i',j')$ similar if $T[i\dd j] = T[i'\dd j']$, except at the 
extremes: let $\alpha_k = \lceil 8 \ell_k \rceil$, then 
$\mathcal{B}_k(i+2\alpha_k,j-\alpha_k) = \mathcal{B}_k(i'+2\alpha_k,j'-\alpha_k)$.

An additional property of the resulting grammar is that it is {\em locally
balanced}: the subtree of the parse tree rooted at nonterminal $A$ is of height 
$O(\log |exp(A)|)$. This is a consequence of the fact that in $S_k$ there are 
fewer than $1+ 4(j-i+1)/\ell_{k+1}$ blocks ending inside $T[i\dd j]$, and the
height of $A$ is never more than the level $k$ of the string $S_k$ where it was
created.

\subsection{Pattern searching}

Let us now define which cuts of $P$ we need to try out in order to capture all
the primary occurrences with this grammar \citep[Sec.~4]{KNOlatin22}. 
Since ends of blocks in $\mathcal{B}_k(i,j)$ correspond to the phrase endings 
where a primary occurrence $T[i\dd j] = P$ can be cut, our set of cutting 
positions must suffice to capture those possible block endings for all $k$ and 
for every possible $T[i\dd j]$ that matches $P$. We define
\begin{eqnarray*}
M_k(i,j) & = & \mathcal{B}_k(i,j) \setminus [2\alpha_{k+1}+1 \dd j-i-\alpha_{k+1}] \\
         & \cup & \{ \min (\mathcal{B}_k(i,j) \cap [2\alpha_{k+1}+1 \dd j-i-\alpha_{k+1}]) \}, 
\end{eqnarray*}
that is, all the cutting points in the extremes, where the different occurrences
of $T[i\dd j]$ may differ, and just the leftmost one in the part that is guaranteed
to be equal. Over all the levels, we build
$$ M(i,j) ~=~ \bigcup_{k \ge 0} M_k(i,j). $$
The key point \citep{KNOlatin22} is that $M(i,j)$ depends only on the content
of $T[i\dd j]$ (not on its position in $T$),
so we can define $M(P) = M(i,j)$ if $P = T[i\dd j]$, and this is the same set
for every possible occurrence of $P$ in $T$. Further, $|M_k(i,j)|=O(1)$ and
$|M(P)| = O(\log m)$.
In operational terms, this means that, at query time, we parse $P$ in $O(m)$
time using the same rules we defined for $T$, producing a parse tree of height
$O(\log m)$ and finding the $O(\log m)$ cutting points $M(P)$.

\section{Faster MEM Finding using LCGs}
\label{sec:fasterMEM}

The idea to use the index of the preceding section is to exploit
the fact that $O(\log (j-i+1))$ cutting points suffice to find all the primary
occurrences of any window $P[i\dd j]$. We will then maintain the parse tree of
$P[i\dd j]$, and the set $M(P[i\dd j])$, as we slide the window through $P$, 
and use them to maintain the number $|R|$ of active positions within 
$O(\log m)$. We also need more sophisticated mechanisms to avoid the quadratic 
costs in lines 6, 7, and 13 of Algorithm~\ref{alg:gmems}.

\begin{algorithm}[t]
        Parse $P$, computing all $P_k$ ; (Section~\ref{sec:parsing})

        Compute nodes $vx_r \in P_{\cal X}$ and $vy_r \in P_{\cal Y}$ for all
$1 \le r < m$ ; (Section~\ref{sec:patricia})

	$i \gets 1$ ; $R \gets \emptyset$ ; 

        \For{$j \gets 0,\ldots,m-1$}
           { \lIf{$i=j+1$ \rm and the root of $P_{\cal X}$ has no child labeled $P[j+1]$}
		{ $i \gets i+1$ }
	     \Else
	        { Advance all $i_k/j_k$ until $P_k[i_k]/P_k[j_k]$ contain
$P[i]/P[j+1]$ ; (Section~\ref{sec:parsing})

	 	  Compute $M(i,j+1)$ ; (Section~\ref{sec:cutting}) 

		  \For{$r \in R$}
		      { \lIf{$r \not\in M(i,j+1)$ {\bf or} $r+|vy_r| \le j$}
			   { $R \gets R \setminus \{r\}$ }
			\lElse
			   { $y_r \gets \textrm{child of } y_r
				  \textrm{ by } P[j+1]$ }
		      }

		  \For{$r \in M(i,j+1) \setminus R$
{\rm (Section~\ref{sec:cutting})}} 
		      { \If{$r+|vy_r| > j$}
			   { $R \gets R \cup \{r\}$ ;

			     $y_r \gets \mathrm{weightedAncestor}(P_{\cal
Y},vy_r,j-r+1)$ ;
			   }
		      }

		  \For{$r \in R$}
		      { $x_r \gets \mathrm{rangeExpand}([vx_r^1,vx_r^2],[y_r^1,y_r^2])$ ;
(Section~\ref{sec:emptiness})

			\lIf{$|x_r|=0$} { $R \leftarrow R \setminus \{r\}$ }
		      }

		  $r_{\min} \gets \mathrm{argmin} \{ r-|x_r|,~ r \in R\}$ ;

		  $l \gets r_{\min}-|x_{r_{\min}}|+1$ ;

		  \If{$l>i$}
		     { \textbf{report} $(i,j)$ with position $T[p-j+i\dd p]$ ;

		       $i \gets l$ ;
		     }
		  \If{$i \le j+1$}
		     { $p \gets j+1-r+\textrm{some text position in } 
				[x_{r_{\min}}^1,x_{r_{\min}}^2] \times [y_{r_{\min}}^1,y_{r_{\min}}^2]$
		     }
	        }
            }
	 \lIf{$i \le m$} {\textbf{report} $(i,m)$ with position $T[p-m+i\dd p]$ }
\vspace{5mm}
\caption{Finding the MEMs of $P[1\dd m]$ in $T$ using our LCG-based index.}
\label{alg:lcgmems}
\end{algorithm}

The new algorithm is shown in Algorithm~\ref{alg:lcgmems}. We first parse the
pattern in a way analogous to the text in line 1, in $O(m\log\log m)$ time. We
then compute in line 2 the deepest nodes $vx_r$ and $vy_r$ we can reach in 
$P_{\cal X}$ and $P_{\cal Y}$ by descending with $P[\dd r]^{rev}$ and 
$P[r+1\dd]$, respectively; note the depth $\ell_r$ of Algorithm~\ref{alg:gmems}
is now $|vy_r|$. These deepest nodes are computed for all $1 \le r < m$ at once 
because this allows a more efficient computation, $O(m\log^2 m)$ time instead
of quadratic. 

In the main loop, lines 7--8 compute the set $M(i,j+1) = 
M(P[i\dd j+1])$ of $O(\log m)$ cutting points for $P[i\dd j+1]$, in 
$O(m\log m)$ overall time.

Lines 9--22 update $R$ according to this new set. Lines 9--12 remove the
active positions $r$ that do not anymore belong to $M(i,j+1)$ or do not
reach $P[j+1]$. The node $y_r$ for the remaining positions is updated by
descending in $P_{\cal Y}$; as in Algorithm~\ref{alg:gmems}, this does not
need access to the text because the path towards $vy_r$ has been verified. 
This takes $O(m\log m)$ time. 

Lines 13--17 
incorporate new active positions in $R$, which appeared in $M(i,j+1)$ and
reach $P[j+1]$; now a position $r$ can enter and leave $M(i,j)$, and hence
$R$, more than once. Further, unlike in Algorithm~\ref{alg:gmems}, $r$ may
enter $R$ not only when $r=j+1$, so $y_r$ is not necessarily the root of 
$P_{\cal Y}$. Instead, $y_r$ is now found as the highest ancestor of $vy_r$ 
in $P_{\cal Y}$ with 
string depth $\ge j-r+1$. This corresponds to a {\em weighted ancestor query}, 
where the weights of the nodes $y$ are their string lengths $|y|$. This query 
can be solved in time $O(\log\log n)$, within $O(|P_{\cal Y}|)$ space and 
preprocessing time \citep{ALLS07}. As we see in Section~\ref{sec:cutting}, this
loop executes $O(m)$ times overall, so it contributes 
$O(m\log\log n)$ to the total time. 

Finally, since all the ranges $[y_r^1,y_r^2]$ have potentially changed by
including $P[j+1]$, lines 19--22 recompute for every $r \in R$ the lowest 
ancestor $x_r$ of the deepest node $vx_r$ such that $[x_r^1,x_r^2] \times
[y_r^1,y_r^2]$ is nonempty. This is not done by successively going to the
parent in $P_{\cal X}$ as in Algorithm~\ref{alg:gmems}, but using so-called
(orthogonal) range successor queries. The total time of these queries along the
whole process is $O(m\log m \log^\epsilon g)$, where $\epsilon>0$ is any
constant. The positions $r$ such that their node $x_r$ becomes the root of
$P_{\cal X}$ are removed from $R$.

Lines 23--34 are as in Algorithm~\ref{alg:gmems}; again our range successor
queries yield some point within the nonempty rectangle they find, for line 24.
The total query time is then $O(m(\log^2 m + \log\log n + \log m\log^\epsilon
g)) \subseteq O(m\log m(\log m + \log^\epsilon n))$. 
We now describe all the algorithmic components of our solution.

\subsection{Parsing the pattern}
\label{sec:parsing}

In this section we preprocess the pattern so that, later, we can
efficiently have the parse tree of every substring $P[i\dd j]$ as we slide the
window along $P$. 
We first parse the whole $P$, using the same permutations $\pi_k$ used to
parse $T$, so as to create sequences $P_k$ analogous to those $S_k$ built when 
parsing $T$. We note that $|P_k| = O(m/\ell_k)$ 
\citep[Lem.~3.5]{KNOlatin22}, thus the number of levels $k$ is $O(\log m)$ 
and the total size of the parse is $O(m)$.
\citet[Sec.~5.2]{KNOlatin22}, modifying the method described by
\citet[Sec.~6.2]{CEKNPtalg20}, show how to obtain all the sequences $P_k$ in
$O(m)$ time: they start from $P_0 = \#P\$$, where $\#$ and $\$$ are symbols
that do not appear in $T$. In every level, apart from using the runs and
local minima rules to define the blocks (Section~\ref{sec:lcg}), they use 
perfect hash functions to find the nonterminals assigned to those blocks 
when they appeared in $T$. Those hash functions are built at indexing time.

As the window $P[i\dd j]$ slides over $P$, we will maintain indices $i_k$ and 
$j_k$ for all $k$, so that $P_k[i_k\dd j_k]$ is the minimal window of $P_k$ 
covering $P[i\dd j]$ (for $k=0$ this is $i_0=i+1$ and $j_0=j+1$). When $j$ or 
$i$ increase, we check whether each $j_k$ and $i_k$ must be increased as well
(so we maintain $|exp(A)|$ for every nonterminal $A$).

While the total time to maintain the parse is clearly in $O(m\log m)$,
it can also be shown to be $O(m)$. The reason is that, when $j$ (resp., $i$)
increases, we verify from $k=0$ upwards if $j_k$ (resp., $i_k$) needs to
increase in order to minimally cover $P[i\dd j]$. Because all the cuts in 
$P_{k+1}$ are also in $P_k$, we can stop as soon as $j_k$ (resp., $i_k$) does 
not increase, without need to verify the higher values of $k$. Therefore, each 
unit of work in $P_{k+1}$ can be charged to a new reached block $P_k[j_k]$ 
(resp., to a just abandoned block $P_k[i_k]$). All the work then adds up to 
$O(m)$, the total length of all the strings $P_k$.

Note that, even if $P$ appears in $T$, the parsing of $P$ may generate blocks
that do not occur in $T$, and thus we will not have nonterminals for those in
the grammar of $T$. We assign arbitrary unused permutation values $\pi_k$
to those blocks (e.g., larger than all the existing ones) to those new blocks.
Blocks in higher levels containing those new symbols will be necessarily new 
as well. As explained in Section~\ref{sec:lcg}, the generated grammar is still 
of size $O(m)$ and height $O(\log m)$. 
Further, the search algorithm stays correct, because the new 
blocks we find and the symbols we assign them form a valid parse of the text
$T'=T\$^n\#P\$$: in $S_1$ we would have formed the run $\$^n$, whose length
would have paused it until the end of the process, so $\#P\$$ would be parsed 
separately of $T$. If our parse of $\#P\$$ missed an occurrence in $T$, then
it would also miss an occurrence in $T'$, where all the blocks found in the
parsing of the pattern are known.

We need dynamic data structures to maintain those new symbols we produce, so
as to recognize them if they reappear in $P_k$. We must store keys $(a,l)$ 
associated with runs $a^\ell$ of odd levels, and tries for the blocks of the 
even levels. Since both the values $(a,l)$ and the number of children in trie
nodes belong to discrete universes of size $O(m)$, we can use dynamic
predecessor data structures \citep{PT06} to operate in time $O(\log\log m)$
in all cases. Added over the $O(m)$ symbols we process along the parse,
the time to create and search unknown blocks is bounded by $O(m\log\log m)$.

\subsection{Maintaining the cutting points and $R$}
\label{sec:cutting}

We compute $M(P[i\dd j])$ as follows. Per its definition, $M_k(i,j)$ 
contains the positions in $P$ of blocks ending within $P_k[i_k\dd j_k-1]$ that 
(i) belong to $[i_0\dd i_0+2\alpha_{k+1}-1]$, (ii) belong to 
$[j_0-\alpha_{k+1}\dd j_0-1]$, or (iii)
are the leftmost in $[i_0+2\alpha_{k+1}\dd j-1]$. Those are collected
consecutively to the right of $P_k[i_k]$ and to the left of $P_k[j_k]$.
Further, there are a constant number of such block endings in every level $k$,
because a segment of length $O(\alpha_{k+1})$ contains only $O(1)$ block endings
in $P_k$ \citep[Lem.~3.8]{KNOlatin22}.

\begin{figure}[t]
\centering
\includegraphics[width=0.7\textwidth]{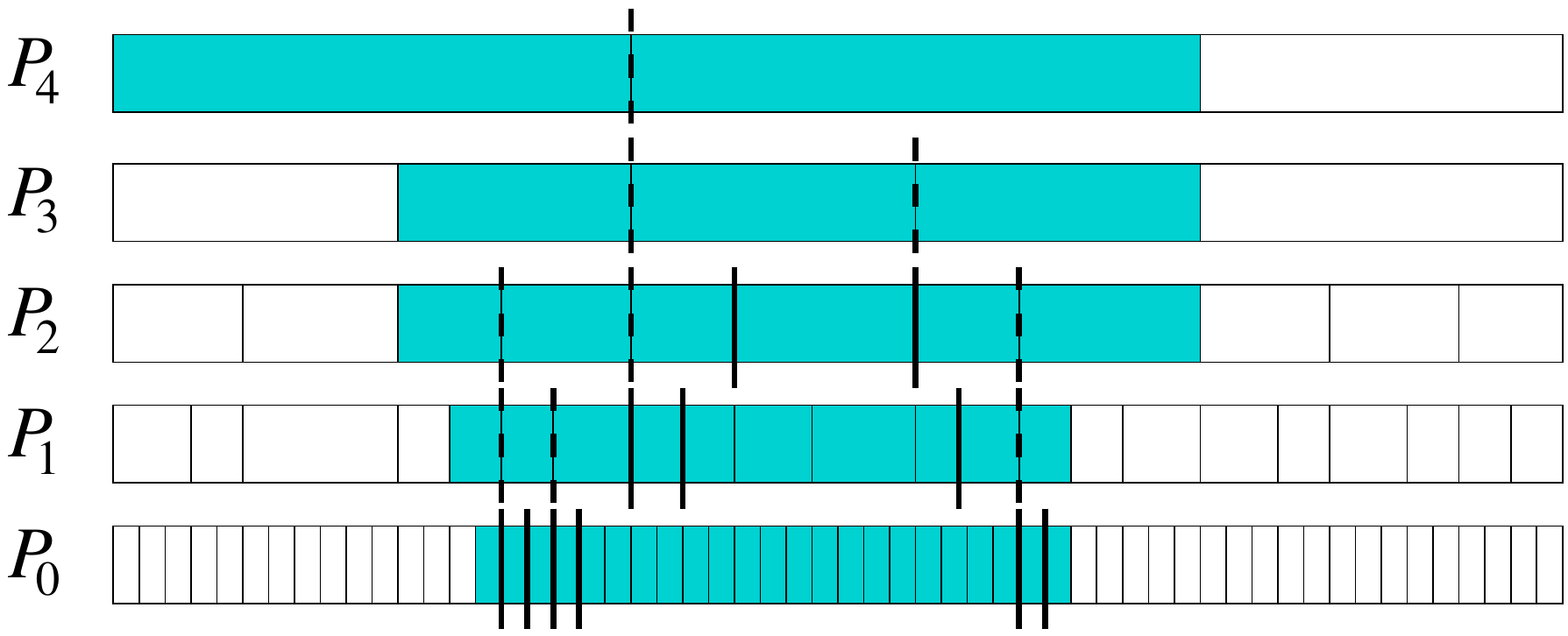}
\caption{A schematic view of a parsing of $P$, shading the involved ranges 
$P_k[i_k\dd j_k]$, and the block ends that make it to $M_k(i,j)$, assuming they
happen to be the first 4 and the last 2 in each level. Dashed segments
indicate block ends that are already reported in a lower level.}
\label{fig:parse}
\end{figure}

We merge all the sets $M_k(i,j)$ to form $M(i,j)=M(P[i\dd j])$. This
set is easily created in sorted order and removing duplicates, because all the 
block ends of $P_{k+1}[i_{k+1}\dd j_{k+1}-1]$ also exist in 
$P_k[i_k\dd j_k-1]$. Therefore for (i) we must collect the block ends of 
$P_0[i_0\dd j_0-1]$ covering $[i_0\dd i_0+2\alpha_1-1]$, then append the block
ends of $P_1[i_1\dd j_1-1]$ covering $[i_0+2\alpha_1\dd i_0+2\alpha_2-1]$, 
then the blocks ends of $P_2[i_2\dd j_2-1]$ covering $[i_0+2\alpha_2\dd 
i_0+2\alpha_3-1]$, and so on. The reasoning is analogous for (ii), prepending 
for increasing $k$ the block ends of $P_k[i_k\dd j_k-1]$ 
that are in $[j_0-\alpha_{k+1}+1\dd j_0-\alpha_k]$. Finally, the sequence
(iii) of the leftmost block ends in $[i_0+2\alpha_{k+1}\dd]$ is increasing with $k$,
so it can be merged with (i) and (ii) in time $O(\log m)$. 
See Fig.~\ref{fig:parse}.

Just as in Section~\ref{sec:parsing}, the total time of this computation can
be made $O(m)$ by working only as long as $j_k$ (resp., $i_k$) increases, so
that the constant amount of work around $i_{k+1}$ and $j_{k+1}$ can be charged 
to the linear traversal of the window $P_k[i_k\dd j_k]$. Precisely, if 
$i_0+2\alpha_{k+1}-1$ (resp., $j_0-\alpha_{k+1}-1$) does not coincide with 
a block end in $P_k$, then no further changes will occur in $M_{k'}(i,j)$ for
any $k' \ge k$. Further, the changes that occur in the lists 
$M_{k'}(P[i\dd j])$ for $k' < k$ are placed within the $O(k)$ first and last
positions in $M(i,j)$, so these updates can be done in $O(1)$ amortized time
per increase of $j$ or $i$. The number of times the loop of lines 13--18 
executes, consequently, is also $O(m)$ along the whole process.

\subsection{Patricia tree searches}
\label{sec:patricia}

As explained, we will find beforehand the deepest nodes $vx_r \in P_{\cal X}$
for all $P[\dd r]^{rev}$ and $vy_r \in P_{\cal Y}$ for all $P[r+1\dd]$.
We make use of the following result,
which was key to obtain subquadratic times in grammar-based indexing. 

\medskip

\begin{lemma}[{\citet[Lem.~6.5]{CEKNPtalg20}}] \label{lem:ztrie}
Let $\cal S$ be a set of strings and assume we have a data structure supporting
extraction of any length-$\ell$ prefix of strings in $\cal S$ in time 
$f_e(\ell)$ and computation of a given Karp-Rabin signature $\kappa$ of any 
length-$\ell$ prefix of strings in {\cal S} in time $f_h(\ell)$. We can 
then build a data structure of size $O(|{\cal S}|)$ such that, later, given
a pattern $P[1 \dd m]$ and $\tau$ suffixes $Q_1,\ldots,Q_\tau$ of $P$, we 
find the ranges of strings in (the lexicographically-sorted) set ${\cal S}$ 
prefixed by each $Q_i$, in $O(m+\tau(f_h(m)+\log m)+f_e(m))$ total time.
\end{lemma}

\medskip

The Karp-Rabin function $\kappa$ \citep{KR87} used in the lemma guarantees no 
collisions between substrings of $T$, so the resulting searches are always 
correct for the suffixes $Q_i$ that occur as a prefix in $\cal S$; for the 
others the structure correctly answers $Q_i$ does not occur. 

When ${\cal S} = \cal X$ or ${\cal S} = \cal Y$, our data structure 
$\cal A$ provides the required prefix/suffix extraction in time $f_e(\ell)=
O(\ell)$. We consider next how to compute the signatures $\kappa$.
We focus on the more complicated case of $Y \in \cal Y$; the 
case of $X \in \cal X$ is analogous. 

\subsubsection{Computing signatures}

A result of independent interest is that we can obtain $f_h(\ell) = O(\log\ell)$
time on our grammar. 
Recall that, on Karp-Rabin signatures.  we can compute
in $O(1)$ time one of $\kappa(S \cdot S')$, $\kappa(S)$, and $\kappa(S')$,
given the other two \citep[Sec.~A.3]{CEKNPtalg20}.

\medskip

\begin{lemma} \label{lem:kr}
The Karp-Rabin signature $\kappa(Y[\dd \ell])$ of any $Y \in \cal Y$ can
be computed in time $O(\log\ell)$ with our grammar.
\end{lemma}
\begin{proof}
We build on the same 
structure $\cal A$ used for extraction from the root of $P_{\cal Y}$.
The strings in $\cal Y$ are concatenations $Y = exp(B_s)\cdots exp(B_t)$
of siblings in rules $A \rightarrow B_1 \cdots B_t$ of the grammar tree. 
The node $v \in P_{\cal Y}$ of $Y$ stores $\langle v \rangle = B_s$.
Let us first assume that $|exp(B_s)| \ge \ell$, so the 
signature can be computed on $exp(B_s)[\dd \ell]$. 

Structure $\cal A$ \citep{GKPS05} is a set of tries on the grammar symbols. 
The terminals 
$\Sigma$ form the trie roots. If $A \rightarrow B_1\cdots B_t$, then $B_1$ is 
the parent of $A$. If $A \rightarrow B^t$, then $B$ is the parent of $A$. Any 
ancestors $C,D$ of $B_s$ in the tries are nodes that descend from $B_s$ by the 
leftmost path in the parse tree. The structure $\cal A$ can jump from $B_s$ to 
any trie ancestor in constant time. Our grammar is locally balanced:
there can be only one block ending inside $exp(B_s[\dd\ell])$ at levels
$k \ge 1+2\log_{4/3} (4\ell)$ \citep[Lem.~3.8]{KNOlatin22}. Thus, the parent
node $C$ of the nonterminal $D$ formed at level $k$ has height $d \le k+1$
in the parse tree and its expansion $exp(C)$ contains $exp(B_s[\dd \ell])$
as a prefix. The node $C$ can then be found in $O(\log\log \ell)$ time 
with exponential search on the ancestors of $B_s$ in the trie.
Therefore, we have that $exp(B_s)[\dd \ell] = 
exp(C)[\dd \ell]$ and can compute the signature on $C$ instead. 
See Fig.~\ref{fig:leszek}.

\begin{figure}[t]
\centering
\includegraphics[width=\textwidth]{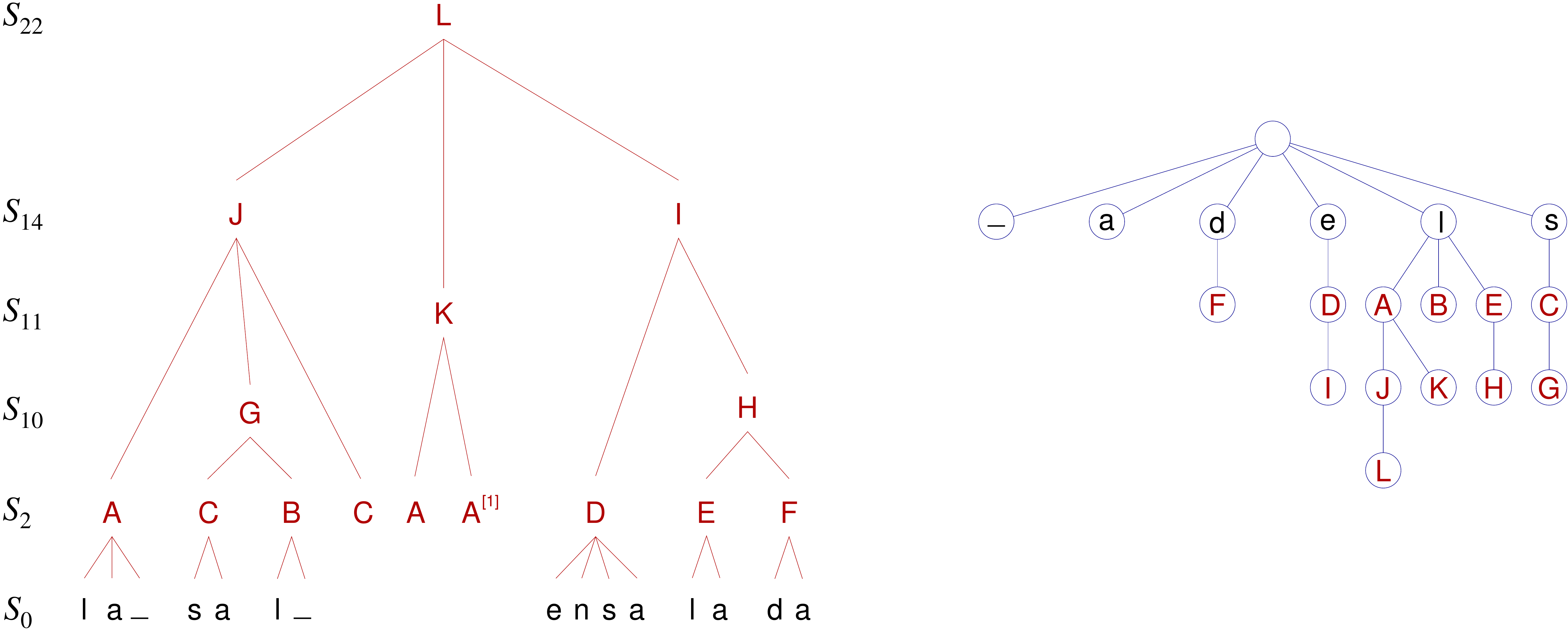}
\caption{On the left, the grammar tree of the LCG of Fig.~\ref{fig:parse},
distinguishing the levels where the nodes are formed. On the right, the data
structure $\cal A$ for prefix access. To compute $\kappa(exp(L)[\dd 6]) = 
\kappa(\textsf{``la\_sal"})$, we search the ancestors of \textsf{L} in $\cal A$
for the lowest one with expansion length at least $6$, which is 
\textsf{J}. From \textsf{J} we go towards the 6th leaf, doing the downward
path $\textsf{J} \rightarrow \textsf{G} \rightarrow \textsf{B} \rightarrow
\textsf{l}$ in the grammar tree, composing the signatures of the left siblings 
in the path: $\kappa(exp(\textsf{A}))$, $\kappa(exp(\textsf{C}))$, and 
$\kappa(exp(\textsf{l}))$.}
\label{fig:leszek}
\end{figure}

The basic algorithm to compute signatures takes time $O(\log^2 \ell)$
\citep[Lem~6.7]{CEKNPtalg20}. It moves from $C$ towards the leaf $L$ of the 
parse tree that corresponds to $exp(C)[\ell]$. Let $C \rightarrow C_1 \cdots
C_t$, then it stores every $w_i(C)=|exp(C_1\cdots C_i)|$ and every
$\kappa_i(C) = \kappa(exp(C_1\cdots C_i))$. The algorithm finds, in 
$O(\log i) \subseteq O(\log\ell)$ time, using exponential search, the $C_i$ that
is in the path to $L$ (i.e., $w_{i-1} < \ell \le w_i$), sets 
$\ell \gets \ell-w_{i-1}$, collects $\kappa_{i-1}(C)$, and continues by $C_i$. 
It composes all those
$\kappa$ values towards $L$ to obtain $\kappa(Y[\dd\ell])$. In rules 
$C \rightarrow C_1^t$ it obtains $i$ in constant time but spends $O(\log i)$ 
time to compute $\kappa_{i-1}(C)$ from the stored $\kappa(exp(C_1))$.

Instead, an $O(\log n)$ time algorithm \citep[Thm.~A.3]{CEKNPtalg20}%
\footnote{They build on the scheme of \citet{BGCSVV17} and
mention in passing that they could compute the fingerprint in $O(\log\ell)$
time with their grammar, but they had overlooked several issues we fix only
now.} replaces
the exponential searches by a more sophisticated scheme, whose cost is the 
telescoping sum $\sum_{h=1}^p \log(t_h/t_{h-1}) \le \log t_p$, where $t_h$ is 
the number of children (counting $C \rightarrow C_1^t$ as having $t$ children) 
of the ancestor at distance $h$ of leaf $L$. In their case, they start from the 
root, which could have $t_p=n$, but if we start it from a node $C$, its time is
$\log t_p \le \log |exp(C)|$. Another component of the cost is the number of 
times one leaves from heavy paths; this is again $O(\log n)$ in general but 
just $O(d)=O(\log\ell)$ if we start from the position of $C$ in its heavy path.

It could be, however, that $exp(C)$ is as long as $n$. Because it was formed
in $S_k$, however, the child $D = C_i$ of $C$ belongs to ${\cal A}_k$ (only 
those nonterminals are allowed to form rules in $S_k$), and thus by definition 
$|exp(C_i)| \le \ell_k$ and $\log |exp(C_i)| = O(k) = O(\log\ell)$. We can then
find $i$ and compute $\kappa_{i-1}(C)$ in time $O(\log i) \subseteq O(\log\ell)$
with the basic method \citep[Lem~6.7]{CEKNPtalg20} and then continue from $C_i$,
where the more sophisticated technique \citep[Thm.~A.3]{CEKNPtalg20} completes 
the computation in another $O(\log |exp(C_i)|) \subseteq O(\log\ell)$ time.

In case $|exp(B_s)| < \ell$, we find the first $s < i \le t$ such that 
$w_i(A) \ge \ell$, and compute instead the signature of $exp(B_i)[\dd \ell-
w_{i-1}(A)]$, to then compose it with the stored values $\kappa_{s-1}(A)$ and
$\kappa_{i-1}(A)$ to obtain the final signature $\kappa(Y[\dd\ell]) =
\kappa(exp(A)[w_{s-1}(A)+1\dd w_{s-1}(A)+\ell])$. 
\end{proof}

\subsubsection{Batched searches}

The search for all the nodes $vy_r$, $1\le r < m$, corresponds to 
searching $P_{\cal Y}$ for every suffix $P[r+1\dd]$ and returning the deepest
reached node. Note that 
Lemma~\ref{lem:ztrie} does not yield the node $vy_r$, but rather its 
corresponding range $[vy_r^1,vy_r^2]$. By performing a lowest common ancestor 
(LCA) query on $P_{\cal Y}$ from the $vy_r^1$th and $vy_r^2$th leaves, we obtain
$v=lca(vy_r^1,vy_r^2)$ (identifying leaves with their ranks). The answer is 
indeed $vy_r = v$ if $|v| = m-r$; if $m-r < |v|$ then $vy_r$ is the virtual 
node of string length $m-r$ on the edge of $P_{\cal Y}$ that leads to $v$.
Linear-space LCA data structures that are built in linear time and answer $lca$
in $O(1)$ time are well known \citep{BFCPSS05}. 

It seems Lemma~\ref{lem:ztrie} will perform the desired searches in time
$O(m\log m)$. The problem is that the lemma works only if $P[r+1\dd]$ actually 
prefixes some string in $\cal Y$. Otherwise, unlike classical trie searching,
it does not yield the maximum prefix of $P[r+1\dd]$ that prefixes some 
string in $\cal Y$. We will resort to, essentially, binary searching for those
longest prefixes using Lemma~\ref{lem:ztrie} as an internal tool.

\begin{figure}[t]
\centering
\includegraphics[width=0.6\textwidth]{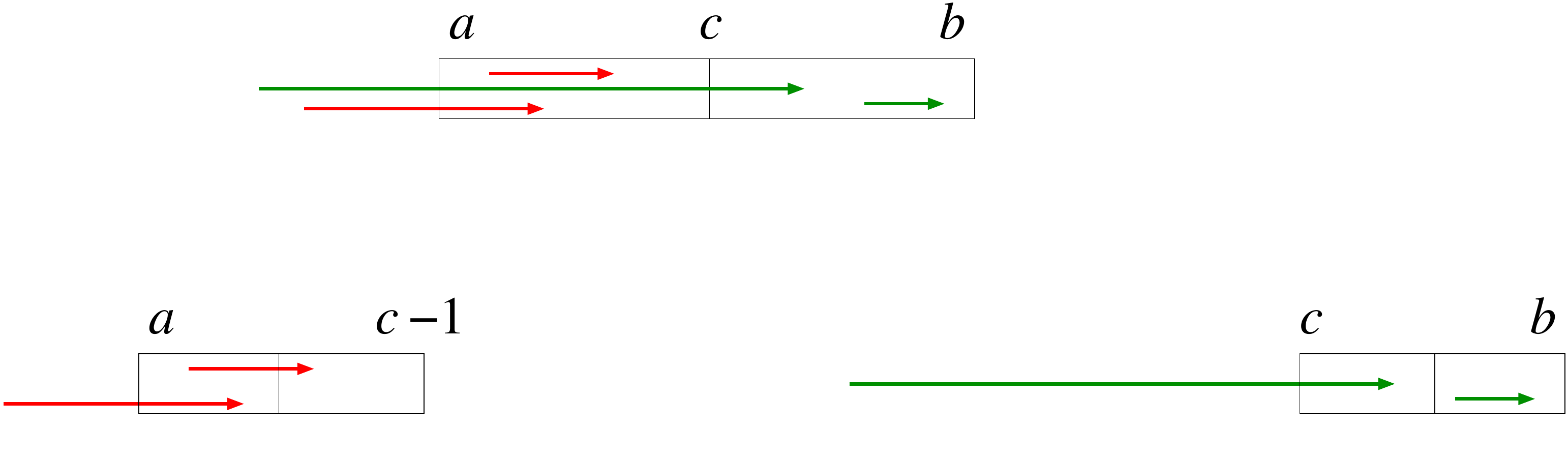}
\caption{Illustration of the suffixes that belong to $\mathcal{Q}_{a,b}$ (top).
The algorithm only knows that the prefix they share with $\cal Y$ ends at
or after $a$ but does not exceed $b$. After testing whether they reach $c$, 
the suffixes are moved either to $\mathcal{Q}_{a,c-1}$ (bottom left) or to 
$\mathcal{Q}_{c,b}$ (bottom right), and tested again.}
\label{fig:Q}
\end{figure}

Assume $m$ is a power of $2$ for simplicity; the general case is easily deduced.
We define sets ${\cal Q}_{a,b}$ of positions, containing those values $r$ such 
that $P[r+1\dd a]$ is known to prefix a string in $\cal Y$ and $P[r+1\dd b+1]$ 
is known not to prefix a string in $\cal Y$ (if $r+1 > a$, then $P[r+1\dd a]
=\varepsilon$ prefixes all strings in $\cal Y$). 
We start with the set ${\cal Q}_{1,m} = \{1,\ldots,m\}$.
To process a set ${\cal Q}_{a,b}$, we search for all the $\tau=|{\cal Q}_{a,b}|$
suffixes $\{ P[r+1\dd c],~r \in {\cal Q}_{a,b} \}$ of $P[\dd c]$ using 
Lemma~\ref{lem:ztrie}, with $c = (a+b+1)/2$. The values $r$ where $P[r+1\dd c]$
is found are moved to ${\cal Q}_{c,b}$, and the others to ${\cal Q}_{a,c-1}$ 
(if $r+1 > c$, we directly move $r$ to ${\cal Q}_{c,b}$ without searching for 
it). Whenever $P[r+1\dd c]$ is
found in $\cal Y$, we will associate with $r$ the corresponding node 
$v_{r,c} \in P_{\cal Y}$; instead, we will retain their previous node 
$v_{r,a}$ when $P[r+1\dd c]$ is not found. 
(In the beginning all such nodes are $v_{r,r}$ and 
equal the root of $P_{\cal Y}$.) See Fig.~\ref{fig:Q}

Note that the values $b-a+1$ halve as the elements in ${\cal Q}_{a,b}$ are
separated into two sets. Any value $r$ is then moved $O(\log m)$ times until 
it ends up in a set of the form ${\cal Q}_{c,c}$; at this point we know that 
the longest prefix of $P[r+1\dd]$ that is also a prefix in $\cal Y$ is 
$P[r+1\dd c]$, and also know its node $v_{r,c} = vy_r$.

The cost of using Lemma~\ref{lem:ztrie} has two parts. The cost $f_h(m)+
\log m = O(\log m)$
can be charged to each of the $\tau$ suffixes sought, and there is an 
additional global cost of $m+f_e(m) = O(m)$. Since every suffix $P[r+1\dd]$
participates $O(\log m)$ times in the lemma, the first cost adds up to
$O(m\log^2 m)$
over all the $m-1$ positions $r$. The second part is potentially very large,
however: the suffixes in ${\cal Q}_{a,b}$ may start well to the left of
$a$, thus the pattern is $P[\dd c]$, not $P[a\dd c]$; a simple application of
the lemma would lead to a quadratic cost again. We address this problem next.

\subsubsection{Smarter substring extractions}

To reduce this time, we consider where the $O(m)$ cost in Lemma~\ref{lem:ztrie}
comes from. A first part refers to the time needed to compute the Karp-Rabin
signatures for all the suffixes in ${\cal Q}_{a,b}$. This cost is easily 
maintained within $O(m)$ overall because we can compute the signatures 
$\kappa(P[r+1\dd])$, for all $1 \le r < m$, in a single pass over $P$, and 
then any $\kappa(P[r+1\dd j])$ is obtained in constant time from 
$\kappa(P[r+1\dd])$ and $\kappa(P[j+1\dd])$.

The second 
part of the $O(m)$ cost corresponds to the time $f_e(m)$ to verify the longest
suffix among those that passed some previous filters; the rest of the 
verification is built on that extracted suffix. Let $P[r+1\dd c]$ be the 
longest candidate suffix. If $r+1 > a$, we extract the actual suffix 
$P[r+1\dd c]$ regularly in time $f_e(c-r) = O(b-a)$ with $\cal A$, because 
$P[r+1\dd]$ starts at the root of $P_{\cal Y}$.

Otherwise, $r+1 \le a$ and thus $P[r+1\dd a]$ had been successfully matched 
before and we have its node $v_{r,a} \in P_{\cal Y}$. As mentioned, the 
process of Lemma~\ref{lem:ztrie} performs several checks before doing the 
final extraction of the longest suffix surviving the checks. We will add a new 
check to those, which can only speed up the process: the candidate node $v$ for 
$P[r+1\dd c]$ must now descend from $v_{r,a}$ in order to be further
considered. The descendance check is performed in constant time by comparing 
the leaf range $[v^1,v^2]$ of $v$ with that of $v_{r,a}$. 
If $v$ passes the test, we know that it does start with 
$P[r+1\dd a]$, and then only need to extract $P[a+1\dd c]$ from the text, 
which is of length $O(b-a)$. 

This time, however, the string to extract does not start at the root of 
$P_{\cal Y}$, and thus it requires a random access to $T$.%
\footnote{It is tempting to say that, since we had already matched 
$P[r+1\dd a]$ from the root of $P_{\cal Y}$, we could somehow save the state of 
that extraction so as to continue without paying the overhead of the random 
access. However, we might have never extracted the text of the node $v_{r,a}$
explicitly; its verification may have been carried out as a subproduct
of reading a longer suffix, starting before $r+1$.}
Say that the node $v$ to verify represents the string 
$Y=exp(B_s)\cdots exp(B_t) \in \cal Y$, of which we want to extract 
$Y[a-r+1\dd c-r]$ to compare it with $P[a+1\dd c]$.
Recall, as in Lemma~\ref{lem:kr}, that
the strings in $\cal Y$ are concatenations $Y = exp(B_s)\cdots exp(B_t)$
of consecutive siblings in rules $A \rightarrow B_1 \cdots B_t$ of the grammar
tree (if $A \rightarrow B^t$, then the node stores $\langle v \rangle =
B^{[t-1]}$ and we have $B_s = B$). 
Let us first assume that $|exp(B_s)| \ge c-r$, so the 
substring to extract is within $exp(B_s)[\dd c-r]$. 
We use again the structure 
$\cal A$, now to extract the string in time $O(b-a+\log c)$.

As in Lemma~\ref{lem:kr}, we can search in $O(\log\log c)$ time 
for the lowest descendant $C$ of $B_s$ such that
$|exp(C)| \ge c-r$; its height is $d=O(\log c)$ because the grammar is locally
balanced. Since $exp(B_s)[\dd c-r] = exp(C)[\dd c-r]$, we descend from $C$
to the leaf $L$ in the parse tree representing $exp(C)[a-r+1]$. Using the same
techniques as in Lemma~\ref{lem:kr}, the time is $O(d) = O(\log c)$. From $L$, 
$exp(C)[a-r+1\dd c-r] = Y[a-r+1\dd c-r]$ is extracted in time
$O(c-a) = O(b-a)$.

In case $|exp(B_s)| < c-r$, the node $C$ is not a descendant of $B_s$ but we
use $C = A$ instead. Given the limitation on $|exp(B_s)|$, the height of 
$B_s$ is $O(\log c)$, and so is the height of $A$. 

The $O(\log c) \subseteq O(\log m)$ cost is absorbed by other $O(\log m)$ 
costs we have already considered. To account for the $O(b-a)$ terms, let us
arrange the sets $\mathcal{Q}_{a,b}$ into {\em levels}; all the sets with 
the same difference $b-a+1$ belong to the same level. Concretely, level
$0$ corresponds to $\mathcal{Q}_{1,m}$, and if $\mathcal{Q}_{a,b}$ is of
level $\ell$, then its halves $\mathcal{Q}_{a,c-1}$ and $\mathcal{Q}_{c,b}$
are of level $\ell+1$. Level $\ell$ is then formed by $2^\ell$ sets whose
ranges $[a,b]$ partition $[1,m]$. Therefore, the sum of the $b-a$ values
over every level yields $m$, and adding over all the $O(\log m)$ levels
we have that the $O(b-a)$ costs add up to $O(m\log m)$.

\subsection{Range successor queries}
\label{sec:emptiness}

Finally, we face the problem of, given a $y$-coordinate range 
$[y_r^1,y_r^2]$ and the $x$-coordinate range $[vx_r^1,vx_r^2]$ of
the deepest node $vx_r \in P_{\cal X}$ reached by $P[\dd r]^{rev}$, find
the lowest ancestor $x_r$ of $vx_r$ such that the range
$[x_r^1,x_r^2] \times [y_r^1,y_r^2]$ in $\cal G$ is nonempty. 

We will solve this query using (orthogonal) {\em range successor queries}: given
a range $[vx_r^1,vx_r^2] \times [y_r^1,y_r^2]$, we can find the largest value 
$x_< \le vx_r^1$ such that $[x_<,vx_r^2] \times [y_r^1,y_r^2]$ contains a point,
and the smallest value $x_> \ge vx_r^2$ such that $[vx_r^1,x_>] \times 
[y_r^1,y_r^2]$ contains a point. Those queries run in $O(\log^\epsilon g)$ 
time on a grid with $g$ points, using an $O(g)$-space data structure, for any 
constant $\epsilon>0$ defined at construction \citep{NNswat12}; its construction
time is $O(g\sqrt{\log g})$ \citep{BP16}.

The lowest ancestor $x_r$ of $vx_r$ containing some point in 
$[x_r^1,x_r^2] \times [y_r^1,y_r^2]$ must then satisfy $x_r^1 \le x_<$ or 
$x_r^2 \ge x_>$.
In the first case, it is $v_1 = lca(x_<,vx_r^2)$; in the second, it is $v_2 =
lca(vx_r^1,x_>)$. Both $v_1$ and $v_2$ are ancestors of $vx_r$, and thus of each
other. The correct node $x_r$ is then the lowest of $v_1$ and $v_2$, which is 

\begin{figure}[t]
\centering
\includegraphics[width=0.5\textwidth]{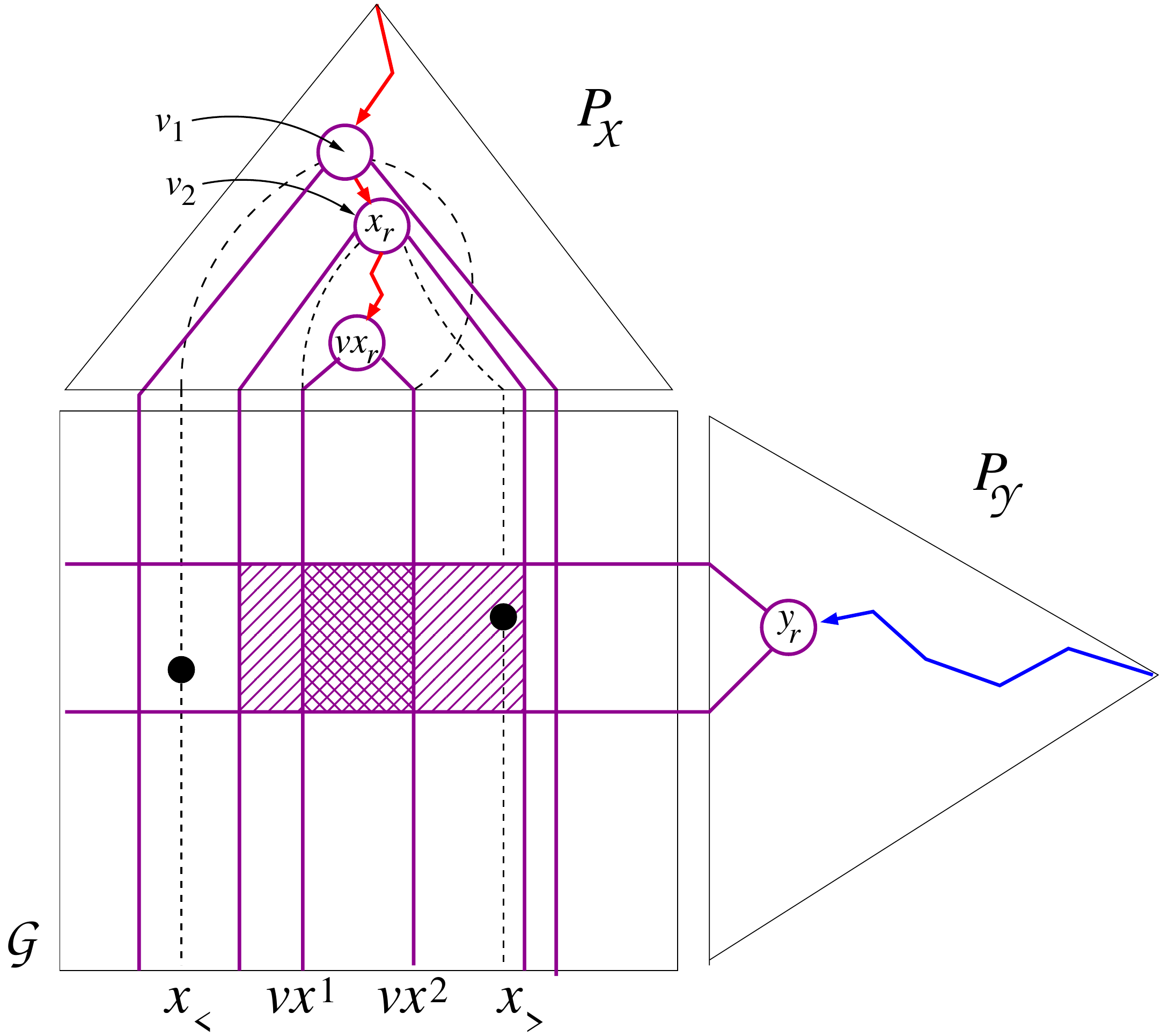}
\caption{Computing the lowest ancestor $x_r$ of $vx_r$ in $P_{\cal X}$ for a
given node $y_r \in P_{\cal Y}$. The initial empty area of $vx_r$ is doubly 
hatched, and it is extended to the hatched nonempty area of $x_r$.}
\label{fig:retreat2}
\end{figure}

\subsection{The final result}

As stated, our time complexities add up to 
$O(m(\log^2 m + \log\log n + \log m\log^\epsilon g_{rl}))$
for a RLCFG of size $g_{rl}$. Since in our case $g_{rl}=
O(\delta\log\frac{n}{\delta})$,
we can write the time as $O(m\log m(\log m + \log^\epsilon\delta+\log\log n))$.
The construction time of all the data 
structures we use is dominated by the $O(n\log n)$ expected time needed 
to build the Karp-Rabin hashes of Lemma~\ref{lem:ztrie} 
\citep[Sec.~6.6]{CEKNPtalg20} (the grammar is built in $O(n)$ expected time, 
see \citet[Cor.~3.15]{KNOlatin22}).

\medskip

\begin{theorem} \label{thm:mem-mlog2m}
Let $T[1\dd n]$ have repetitiveness measure $\delta = \delta(T)$. Then, for 
any constant $\epsilon>0$, we can build a data structure of 
size $O(\delta\log\frac{n}{\delta})$ that finds the MEMs of any given pattern 
$P[1\dd m]$ in time $O(m\log m(\log m + \log^\epsilon \delta+\log\log n))
\subseteq O(m\log m(\log m+\log^\epsilon n))$, returning an occurrence of each 
MEM. The data structure can be built in $O(n\log n)$ expected time.
The query process uses $O(m)$ additional space on top of the data structure.
\end{theorem}

\medskip

We have assumed $m \le n$ throughout, but it could be the other way in some applications.
Since in this case no substring longer than $n$ will be matched inside $P$,
we can run $O(m/n)$ iterations finding the MEMs of $P[1\dd 2n]$, $P[n\dd 3n]$,
$P[2n\dd 4n]$, and so on, avoiding repeated MEMs in the output. The total
cost would then be $O(m\log^2 n)$ and the query space would be $O(n)$.

\section{Related Problems} \label{sec:relatives}

We now apply our findings to solve some problems related to finding MEMs.
In all cases we will still consider that $T$ is indexed with a grammar,
whereas $P$ is given in online form and we can spend $O(m)$ space at query
time.

\subsection{Finding $k$-MEMs} \label{sec:kMEMs}

We first extend our results to finding the $k$-MEMs of $P[1\dd m]$, where $k$
is given at query time together with $P$. The idea is to count the number
of secondary occurrences triggered from the primary occurrences we find in the 
grid. The total number of occurrences of $P[i\dd j]$ in $T$ is the sum of 
those numbers over all the points in the areas $[x_r^1,x_r^2] \times 
[y_r^1,y_r^2]$ corresponding to the active positions $r \in R$.

\begin{algorithm}[t!]

        \While{$l \le j+1$}
	  { $q \gets 0$ ;

            \For{$r \in R$}
	        { \lIf{$l = r-|x_r|+1$}
		     { $q \gets q + \textrm{sum of occurrences in }
                                         [x_r^1,x_r^2] \times [y_r^1,y_r^2]$ }
                }
  	    \lIf{$q \ge k$} { \textbf{break} }
	    \For{$r \in R$}
                 { \If{$l = r-|x_r|+1$}
                     { $x_r \gets \textrm{(virtual) parent of } x_r$ ; 

		       \lIf{$|x_r|=0$} { $R \gets R \setminus \{r\}$ }
		     }
                 }
	  
 	    $l \gets l+1$
	 }

\vspace{5mm}
\caption{Modification to find the $k$-MEMs of $P[1\dd m]$ in $T$ using a 
grammar-based index.}
\label{alg:kmems}
\end{algorithm}

Algorithm~\ref{alg:kmems} shows the lines we must insert right after line 18
of Algorithm~\ref{alg:gmems} to account for $k$-MEMs. In addition, in line 
24 of Algorithm~\ref{alg:gmems} we must collect some of the points reached 
in line 4 of Algorithm~\ref{alg:kmems}, instead of one of $r_{\min}$.

The new invariants for the window $P[i\dd j]$ are that 
(i) we have reported every $k$-MEM ending before $j$, and that either 
(ii.1) $R=\emptyset$ and no $k$-MEM of $P[\dd j]$ ends at $j$, or (ii.2) 
$R \neq \emptyset$ and the $k$-MEM of $P[\dd j]$ ending at $j$ is $(i,j)$. 

The required changes, in Algorithm~\ref{alg:kmems}, are that once we compute
the leftmost position $l$ such that $P[l\dd j+1]$ occurs in $T$, we check in 
lines 2--5 that there are at least $k$ occurrences of $P[l\dd j+1]$. Only the
occurrences of the active positions $r$ that reach position $l$ (i.e., 
$l=r-|x_r|+1$) are counted. If we count less than $k$ occurrences, then we 
must advance $l$, going to the parent of all those nodes $x_r$ that reached 
$l$, and removing the active positions $r$ where $x_r$ becomes the root of 
$P_{\cal X}$.

The dominant term in the time complexity are the queries that count the sum
of occurrences in orthogonal subgrids $[x_r^1,x_r^2] \times [y_r^1,y_r^2]$.
This query is performed $|R|$ times for every new value of $j$ and of $i$ (or
$l$, in our case). 
With the bound $|R| \le m$, we have a total time of $O(\tau m^2)$, where
$\tau$ is the time of the geometric query. 

The counting query can be done by performing up to $k$ orthogonal range 
successor queries on $[x_r^1,x_r^2] \times [y_r^1,y_r^2]$, finding one
primary and zero or more secondary occurrences from each point
we find in the grid. Note that, although the process to find the secondary
occurrences ensures only amortized time \citep{CNPjcss21}, we need to know 
only the number of such occurrences, so we know we have $k$ occurrences as 
soon as we perform $k$ recursive calls along the process of finding the 
secondary occurrences. In general, we stop as soon as we add up to $k$ 
occurrences along all the primary and secondary occurrences triggered by 
all the points found in the consecutive range successor queries on 
$[x_r^1,x_r^2] \times [y_r^1,y_r^2]$. Since
each point brings in at least one primary occurrence, and the range successor 
queries take $O(\log^\epsilon g_{rl})$ time, we obtain 
$\tau = O(k\log^\epsilon g_{rl})$. This yields a 
natural generalization of of Theorem~\ref{thm:mem-m2}.

\medskip

\begin{theorem} \label{thm:kmem-m2}
The data structure of Theorem~\ref{thm:mem-m2} can find the $k$-MEMs of any
given pattern $P[1\dd m]$, for any $k>0$ given with $P$, in time 
$O(k m^2 \log^{\epsilon} g_{rl})$. 
\end{theorem}

\medskip

Algorithm~\ref{alg:kmems} can also be inserted right after line 24 of
Algorithm~\ref{alg:lcgmems}. Now the size of $R$ is bounded by $O(\log m)$,
which makes Algorithm~\ref{alg:kmems} run in time $O(\tau m\log m)$.

\medskip

\begin{theorem} \label{thm:kmem-mlog2m}
The data structure of Theorem~\ref{thm:mem-mlog2m} can find the $k$-MEMs of
any given pattern $P[1\dd m]$, for any $k>0$ given with $P$, in time 
$O(m\log m(\log m + k(\log^\epsilon \delta+\log\log n)))
\subseteq O(m\log m(\log m+k\log^\epsilon n))$.
\end{theorem}

\medskip

These solutions, however, are impractical for large $k$. We can improve the 
time of Theorem~\ref{thm:kmem-m2} by restricting our grammar to be a CFG, of size 
$g \ge g_{rl}$. On CFGs, the number of secondary occurrences triggered by each
primary occurrence is a function of the grid point only, and thus it can be
precomputed and stored with the points. The summing query can then be solved 
directly as a geometric query, in time $O(\log^{2+\epsilon} g)$ for any 
constant $\epsilon>0$ chosen at index construction time, within 
$O(g)$ space \citep[Sec.~6]{Nav18}. The construction time does not change 
with respect to the basic solution. This is then faster than the solution of
Theorem~\ref{thm:kmem-m2} when $k = \omega(\log^2 g)$.

\medskip

\begin{theorem} \label{thm:kmem-m2count}
Assume we have a CFG of size $g$ that generates only $T[1\dd n]$.
Then, for any constant $\epsilon>0$, we can build a data structure of size
$O(g)$ that finds the $k$-MEMs of any given pattern $P[1\dd m]$, for any $k>0$
given with $P$, in time $O(m^2\log^{2+\epsilon} g)$.
The data structure can be built in
$O(g\log^2 n)$ time and $O(g\log n)$ space, or in
$O(n+g\sqrt{\log g})$ time and $O(n)$ space.
The query process uses $O(m)$ additional space on top of the data structure.
\end{theorem}

\medskip

The time can drop to $O(m^2\log g)$ if we use the
$O(g\log g)$ space of a faster and larger geometric representation for $g$
points that answers range counting queries in time $O(\log g)$
\citep{ABR00}. The structure is built in $O(g\log^2 g)$ expected time.

Although it is unknown how to set the point values in the grid 
so that they correctly sum up the secondary occurrences on a general 
RLCFG, the problem can be handled in our 
particular grammar \citep[Sec.~6]{KNOarxiv23} (see also the technique they
build on \citep[Sec.~7]{CEKNPtalg20}, which has more details).
The only obstacle to use the 
technique is that it needs to compute the shortest period $p=p(i,j)$ of any 
window $P[i\dd j]$ we consider (this is the smallest positive value such that 
$P[i\dd j-p] = P[i+p\dd j]$). \citet[Thm.~1.1.2]{Koc18} shows how
to preprocess $P$ in $O(m)$ time so that any $p(i,j)$ can be computed in
$O(\log m)$ time. The total time per counting query is then $\tau = 
O(\log m + \log^{2+\epsilon} g)$ for a grammar of size $g$, and the
query time is $O(\tau m\log m)$.


\medskip

\begin{theorem} \label{thm:kmem-mlog2mcount}
Let $T[1\dd n]$ have repetitiveness measure $\delta = \delta(T)$. Then, for
any constant $\epsilon>0$, we can build a data structure of size 
$O(\delta\log\frac{n}{\delta})$ that finds the $k$-MEMs of any given pattern
$P[1\dd m]$, for any $k>0$ given with $P$, in time 
$O(m\log m(\log m + \log^{2+\epsilon} (\delta\log\frac{n}{\delta})))
\subseteq O(m\log m\log^{2+\epsilon} n)$,
returning an occurrence of each
MEM. The data structure can be built in $O(n\log n)$ expected time.
The query process uses $O(m)$ additional space on top of the data structure.
\end{theorem}

\medskip

Again, by spending $O(\delta\log\frac{n}{\delta}\log n)$ space, the range
counting time drops to $O(\log(\delta\log\frac{n}{\delta}))$
\citep{KNOarxiv23,ABR00}, and thus the total query time becomes
$O(m\log m(\log m + \log (\delta\log\frac{n}{\delta})))
\subseteq O(m\log m\log n)$.

\no{
\begin{theorem} 
Let $T[1\dd n]$ have an attractor of size $\gamma$. Then, for any constant
$\epsilon>0$, we can build in $O(n\log n)$ expected time a data structure of
size $O(\gamma\log\frac{n}{\gamma})$ that finds the $q$-MEMs of any given 
pattern $P[1\dd m]$ and any $q>0$ in time $O(m\log m(\min(\log n,\log^2 m) +
\log^{2+\epsilon} (\gamma\log\frac{n}{\gamma})))
\subseteq O(m\log m\log^{2+\epsilon} n)$.
The query process uses $O(m)$ additional space.
\end{theorem}
}

\subsection{MUMs} \label{sec:mums}

As described in Section~\ref{sec:MEMs}, a MUM between $P$ and $T$ is a maximal 
substring that appears exactly once in both $P$ and $T$. We build on the 
following observation.

\medskip

\begin{lemma}
If $P[i\dd j]$ is a MUM between $P$ and $T$, then it must be a MEM in $T$.
\end{lemma}
\begin{proof}
If $P[i\dd j]$ is a MUM, then it appears exactly once in $P$ and $T$. If it
is not a MEM in $T$, then either $P[i-1\dd j]$ or $P[i\dd j+1]$ appear in 
$T$ as well, and because they cannot be more frequent than $P[i\dd j]$, they appear
once both in $P$ and in $T$. Therefore, $P[i\dd j]$ is not maximal.
\end{proof}

\medskip

Our general strategy is then to find the MEMs of $P$ in $T$ and filter out
those that are not MUMs. We build the suffix tree of $P$ in $O(m)$ time
\citep{FFM00}. Then we run Algorithm~\ref{alg:gmems}$+$\ref{alg:kmems} (i.e.,
adding to Algorithm~\ref{alg:gmems} the lines of Algorithm~\ref{alg:kmems}) 
for finding $k$-MEMs with $k=1$. This seems to be the same as finding MEMs,
but we will modify the $k$-MEM finding algorithm soon.

At the same time, we traverse the suffix tree of $P$, maintaining the locus of 
$P[i\dd j]$ by using the child and suffix link operations, just as described in 
Algorithm~\ref{alg:stmems}. This time, however, we run on the suffix tree of
$P$ itself, not of $T$. Further, we decide whether to increase $j$ (i.e., go
to the child of the current suffix tree node) or to increase $i$ (i.e., go
to the suffix link of the current node) following what
Algorithm~\ref{alg:gmems}$+$\ref{alg:kmems} does. 

We use the implementation of Algorithm~\ref{alg:kmems} used to prove
Theorem~\ref{thm:kmem-m2}. This implementation does not really compute the precise 
number $q$ of occurrences of $P[i\dd j+1]$ in $T$, but it works just enough to 
determine if $q < k$ or $q \ge k$ (for us, $q = 0$ or $q = 1$). This allows it 
to run in time proportional to $k$. We slightly extend its counting process
until determining whether $q=0$, $q=1$, or $q \ge 2$, so that $q=1$ determines 
that $P[i\dd j+1]$ appears exactly once in $T$. We then proceed as for
$k$-MEMs with $k=1$, but in case $q \ge 2$ we record an invalid position in
$p$. 

When it comes to report $(i,j)$ and its occurrence $T[p-j+i\dd p]$, we do so
only if, in addition $(a)$ $p$ is a valid position, which means $P[i\dd j]$
appears only once in $T$, and $(b)$ the locus of $P[i\dd j]$ in its suffix tree 
is (on an edge to) a leaf, which means that $P[i\dd j]$ appears only once in 
$P$. We apply the same filter when reporting $(i,m)$ in the last line of
Algorithm~\ref{alg:gmems}.

In the same way we have combined Algorithms~\ref{alg:gmems} and 
\ref{alg:kmems}, we can combine Algorithms~\ref{alg:lcgmems} and
\ref{alg:kmems}. This yields the following result.

\medskip

\begin{theorem} \label{thm:mum}
The data structures of Theorems~\ref{thm:mem-m2} and \ref{thm:mem-mlog2m}
can find the MUMs of $T$ and any given pattern $P[1\dd m]$, in the same 
asymptotic time they require to find the MEMs. 
The query process uses $O(m)$ additional space.
\end{theorem}

\subsection{Rare MEMs} \label{sec:rare}

A MEM of $P$ in $T$ is $k$-rare if it appears in $P$ and $T$, yet at most $k$
times; MUMs are $1$-rare MEMs. We can generalize the method of 
Section~\ref{sec:mums} so that we still look for $k$-MEMs with $k=1$, but
work enough on the counting stage so as to determine whether $q=0$, 
$1 \le q \le k$, or $q > k$. We store an invalid position $p$ when $q >k$;
otherwise we proceed as for $k=1$. When it comes to report $(i,j)$, we do
so only if $(a)$ position $p$ is valid (i.e., $P[i\dd j]$ appears at most 
$k$ times in $T$) and $(b)$ the locus of $P[i\dd j]$ in the suffix tree of
$v$ is (on the edge to) a node $v$ having at most $k$ descendant leaves
(for which we store the number of leaves that descend from the suffix tree 
nodes, which can be computed in $O(m)$ time once the suffix tree is built).

By counting the occurrences in different ways, we obtain the following result.

\medskip

\begin{theorem} \label{thm:rare}
The data structures of Theorems~\ref{thm:kmem-m2}, \ref{thm:kmem-mlog2m},
\ref{thm:kmem-m2count}, and \ref{thm:kmem-mlog2mcount}
can find the $k$-rare MEMs of $T$ and any given pattern $P[1\dd m]$, in the 
same asymptotic time they require to find the $k$-MEMs of $P$ in $T$. 
The query process uses $O(m)$ additional space.
\end{theorem}

\section{Applications} \label{sec:app}

We now describe other popular problems related to finding MEMs our
results impact.

\subsection{Matching statistics}

A problem directly related to finding MEMs is that of computing the so-called
matching statistics, which we already discussed in the Introduction. While
first defined by \citet{CL94} for approximate string matching, they have a
number of bioinformatic applications like estimating the distance between
genomes for phylogenetic reconstruction, 
estimating cross-entropy, 
computing string kernels, 
string mining, 
and species estimation, 
to name a few \citep{Ohl13,MBCT15}.

\medskip

\begin{definition}
Given a text $T[1\dd n]$ that can be preprocessed, the {\em matching statistics
problem} is that of, given a pattern $P[1\dd m]$, return the length $M[q]$ of 
the longest prefix of $P[q\dd]$ that occurs in $T$, for every $1 \le q \le m$.
A position where each such longest prefix occurs must be given for each $q$.
\end{definition}

\medskip

Given a solution to the MEM finding problem, $(i_1,j_1),\ldots,(i_s,j_s)$,
we compute the matching statistics as follows. Set all $M[q]$ to zero and then 
traverse the tuples $(i_r,j_r)$ in order. Set $M[q] = j_r-q+1$ for 
all $i_r \le q \le \min(j_r,i_{r+1}-1)$, assuming $i_{s+1}=m+1$. 
The occurrence of each $M[q]>0$ is that of its MEM $(i_r,j_r)$ shifted
by $q-i_r$. Conversely, given the matching statistics $M[q]$ for $1\le q\le m$,
we obtain the MEMs by reporting, for increasing $i$, every pair $(i,i+M[i]-1)$ 
such that $i=1$ or $M[i] \ge M[i-1]$, and $M[i]>0$. 

Therefore, both problems are interchangeable as one output can be converted to 
the other in optimal $O(m)$ time. In particular, our results allow computing 
the matching statistics of a patterm $P[1\dd m]$ with respect to a text 
$T[1\dd n]$ that is indexed in compressed form, within the same space and
times seen in Theorems~\ref{thm:mem-m2} and \ref{thm:mem-mlog2m}.

\subsection{MEMs and relatives against a collection of texts}

Some applications, especially related to multiple alignment, consider computing
the MEMs and $k$-MEMs between a pattern $P$ and a collection of texts 
$T_1, \ldots, T_\tau$ \citep[Sec.~7.6]{Gus97}, and even more prominently, the
MUMs \citep[Ch.~11]{MBCT15}.

\medskip

\begin{definition}
The $k$-MEMs between $P$ and a collection of texts $T_1, \ldots, T_\tau$ are
the maximal substrings of $P$ that appear at least $k$ times in each of the
texts $T_t$, for $1 \le t \le \tau$. The case $k=1$ corresponds to the MEMs
between $P$ and $T_1, \ldots, T_\tau$. The $k$-rare MEMs are the
maximal substrings of $P$ that appear in each of the texts $T_t$, for $1 \le t
\le \tau$, but at most $k$ times in each. The case $k=1$ corresponds to MUMs.
\end{definition}

\medskip

While we only consider finding the MEMs and $k$-MEMs between
a pattern $P$ and a text $T$, assume that we have the texts $T_t$ in
compressed form and we compute the list of ($k$-)MEMs 
$M_t = (i^t_1,j^t_1),\ldots,(i^t_{s_t}, j^t_{s_t})$ between $P$ and every 
$T_t$. Those lists come sorted by increasing $i^t$ and by increasing $j^t$. 
We then compute the ($k$-)MEMs between $P$ 
and the collection by running a segment intersection algorithm on the lists
$M_t$: we want the maximal segments that are included in a segment of each
$M_t$. 

We set cursors $c_t = 1$ for all $t$, and compute 
$i = \max \{ i^t_{c_t}, 1 \le t \le \tau \}$ and 
$j = \min \{ j^t_{c_t}, 1 \le t \le \tau \}$. If $i \le j$ we report the
($k$-)MEM $(i,j)$. We then increment the cursor(s) $c_t$ for which $j^t_{c_t}=j$,
and continue until exhausting some of the lists. By using a max-heap
to maintain the values $i^t_{c_t}$ and a min-heap to maintain the 
values $j^t_{c_t}$, the whole process takes time $O(N\log\tau)$, where $N$
is the total length of the lists $M_t$. This complexity is negligible compared
to that of computing the ($k$-)MEMs.

Finding MUMs and $k$-rare MEMs between $P$ and texts $T_1,\ldots,T_\tau$
requires intervening the combined Algorithm~\ref{alg:lcgmems}$+$\ref{alg:kmems}
we developed in Sections~\ref{sec:mums} and \ref{sec:rare}. Concretely, we
run in synchronization the algorithm for every $T_t$, apart from moving in the
suffix tree of $P$. We maintain the same interval $P[i\dd j]$ on all the
texts, skipping the value of $j+1$ if the condition in line 5 of
Algorithm~\ref{alg:lcgmems} is true for any of the texts. Further, the value
$l$ computed in line 24 must correspond to the maximum $l$ over all
the texts. For that value of $l$, we will determine if any of the texts $T_t$
has more than $k$ occurrences of $P[i\dd j]$, and if so, make the text
position invalid so that this window is not reported. The asymptotic cost is
then the same as the sum of the costs of computing MUMs of $k$-rare MEMs
between $P$ and every text $T_t$.

\subsection{Relative Lempel-Ziv compression} \label{sec:rlz}

Relative Lempel-Ziv (RLZ) \citep{KPZ10} is a compression algorithm that first 
chooses a {\em reference} text $R$, and then compresses  texts $T$ by 
representing them as minimum-length sequences of substrings of $R$.
That is, RLZ represents $T$ in $O(z)$ space by choosing a minimal number $z$ 
of substrings of $R$ such that $T = R[a_1\dd b_1] \cdots R[a_z \dd b_z]$.%
\footnote{We assume $R$ contains every alphabet 
symbol, so $T$ can always be covered by substrings of $R$.}
RLZ has proved particularly effective to compress collections where every
text is relatively close to each other, such as in genome collections of
the same species \citep{KPZ10} and web pages \citep{HPZ11}.

The compression algorithm traverses $T[1\dd n]$ left to right, in streaming 
mode. If it has already compressed $T[1\dd i-1]$, it finds the largest $j \in 
[i\dd n]$ such that $T[i\dd j]$ occurs in $R$, say at $R[a\dd b]=T[i\dd j]$, 
and outputs the pair $(a,b)$. Finding those longest prefixes $T[i\dd j]$ of 
$T[i\dd n]$ in $R$ can be 
done in time $O(j-i+1)$ with a suffix tree of $R$, for a total compression
time of $O(n)$. More space-efficient implementations are possible with 
compressed suffix arrays, where the matching of $T[i\dd]$ can proceed 
left-to-right until finding a mismatch. Using a grammar-compressed index for
$R$, instead, is difficult because it is designed to match a fixed $T[i\dd j]$ 
with $j$ known beforehand.

An issue with RLZ is how to choose the reference $R$. In genome collections
of the same species, choosing one random genome is efficient enough, but there
are more complex situations, such as metagenomic collections or other
versioning scenarios. The general premise is to aim for a relatively small
$R$, especially because we need it indexed in main memory in order to 
efficiently compress $T$. Various techniques to choose or generate a good
reference have been studied \citep{KPZacs11,KPZ11,LPMW16,GPV16}. 
On the other hand, however, the larger $R$, the better compression we achieve
in general, because we can potentially choose longer phrases.

With our techniques, we could represent $R$ in grammar-compressed form, thus
$R$ could be a much larger, repetitive, set of texts without compromising the
main memory space used by the index of $R$. 
For example, we could choose the whole set of genomes in
a collection as the reference, so as to compress a new genome $T$ by choosing
pieces from the closest possible individuals. Our MEM-finding technique is
precisely what we need to simulate the searches for prefixes 
$T[i\dd]$ with unknown ending positions. 

To compress $T$, we find the MEMs of $T$ in $R$, $(i_1,j_1),\ldots,(i_s,j_s)$. 
By assuming as before that $R$ contains every alphabet symbol, it holds that 
$i_1=1$, $i_{r+1} \le j_r$ for all $r$, and $j_s = n$. The minimum coverage of 
$T$ then contains $z \le s$ substrings, which can be built from the MEMs in
$O(s)$ time as follows: set cursors at $c=1$ and $p=1$. Output pair $(p,j_c)$. 
Set $p = j_c+1$. Increment $c$ as long as $i_c \le p$. Continue
until $p = n+1$. We can then compress $T$ in time 
$O(n\log n(\log n + \log^\epsilon r))$, where $r=|R|$,
using Theorem~\ref{thm:mem-mlog2m}; the time to cover $T$ with the MEMs is
asymptotically irrelevant.

\subsection{All-pairs suffix-prefix matching}

When assembling a genome from a set of reads without the help of a reference 
genome (a.k.a., de novo fragment assembly), one relevant technique is to form 
a so-called ``overlap graph'', where every node is a read and there is an edge 
of weight $\ell$ from node $u$ to node $v$ iff the $\ell$-length suffix of the 
read of $u$ equals the $\ell$-length prefix of the read of $v$. One then aims, 
in broad terms, to traverse the overlap graph in a way that maximizes the 
weights and collects all the reads \citep[Ch.~4]{SM97}
\citep[Sec.~7.10]{Gus97}. A similar technique is 
used for transcript assembly \citep[Ch.~15]{MBCT15}. To build the overlap graph,
one needs to find all the suffix-prefix matches between every pair of 
reads, disregarding those whose lengths are below a significance threshold
$\ell_{\min}$. This problem, known as the ``all-pairs suffix-prefix matching
problem'', can be solved in optimal time $O(n+e)$, where $n$ is the total 
length of the reads and $e$ is the number of edges in the resulting graph, yet
it needs to build a suffix tree on the concatenation of the reads
\cite[Sec.~8.4.4]{MBCT15} \cite[Sec.~7.10.1]{Gus97}, or an enhanced suffix array
\citep[Sec.~5.6.7]{Ohl13}. If one uses compressed suffix arrays,
then the time complexity becomes $O(n\log\sigma+e)$, where $\sigma$ is the
alphabet size \citep[Sec.~7.7.3]{Ohl13}.

Since collections of reads with high coverage are repetitive, we can use our
LCG of Section~\ref{sec:lcg} to index the whole sequence $T[1\dd n]$ of reads, 
using a special symbol $\$$ as a terminator for each. In order to find all the 
suffix-prefix matches in the set, we process each read $P[1\dd m]$ against $T$.
We run a specialized version of Algorithm~\ref{alg:lcgmems}, only as long as 
$i=1$ (because we are interested in the prefixes of $P$). Since the whole $P$ 
occurs in $T$, however, it is not necessary to check for this condition; we 
will just run the loop of line 4 over lines 7--22, processing each whole read 
$P$.

For each iteration of the loop, where we have processed the window $P[1\dd j]$ 
and now consider $j+1$, we first {\em simulate} that $P[j+1]=\$$, run the lines
7--22, and report {\em all} the occurrences of $P[\dd j] \cdot \$$ in $T$. That
is, we collect all the primary occurrences and also follow all the secondary 
occurrences for all active points $r$ such that $r-|x_r|+1 = 1$ (i.e., they 
match $P$ from position $1$). Each such occurrence corresponds to the suffix 
of some read 
matching the prefix $P[\dd j]$.\footnote{Since $P$ occurs in $T$, there will be 
a spurious match with itself, which is easily filtered out when found.} This
simulation is done only if $j \ge \ell_{\min}$.

After reporting the matches, we {\em undo} the updates performed in lines 7--22
when simulating that $P[j+1]=\$$, and {\em redo} lines 7--22 for the actual 
value of $P[j+1]$. 

Undoing is implemented by just maintaining a copy of the previous set
$R$ and all the values $j_k$, $x_r$, and $y_r$, re-stating them after the
simulation. The time for simulating $P[j+1]=\$$, undoing the simulation,
and then running again with the actual value, adds up to 
$O(\log m \log^\epsilon n)$. Considering the preprocessing of lines 1--2
and adding over all the reads, the total time is in 
$O(n\log m(\log m + \log^\epsilon n))$,
where $m$ is the average length of a read,\footnote{This is because $\log m$
and $\log^2 m$ are convex functions of $m$.} plus the number of prefix-suffix 
matches found. The working space on top of the index is just $O(m)$.

\section{Conclusions} \label{sec:concl}

We have obtained improved results, including the first subquadratic algorithm,
to find MEMs on parsing-based
indices, which are the most promising in terms of space for highly repetitive 
text collections. While suffix-based indices can preprocess $T[1\dd n]$ to find 
the MEMs of $P[1\dd m]$ in $T$ in time $O(m\log\log n)$, their space is 
$\Omega(r)$, where $r$ (the number of runs in the BWT of $T$) is not such a 
strong measure of repetitiveness \citep{Navacmcs20.2}. 
Our first result is a data structure of size $O(g_{rl})$,
where $g_{rl}$ is the size of the smallest RLCFG that generates $T$. This is
currently the best possible space for any structure able to access $T$ with
relevant time guarantees \citep{Navacmcs20.2}. Our structure
finds the MEMs in $O(m^2\log^\epsilon n)$ time for any constant 
$\epsilon > 0$. This is very similar to the time of previous work 
\citep{Gao22}, which could also run in $O(g_{rl})$ space.
Within $O(\delta\log\frac{n}{\delta})$ space, we obtain the first subquadratic 
time, $O(m\log m(\log m +\log^\epsilon n))$, on a particular RLCFG that has 
local consistency properties. This space is optimal for every $n$ and $\delta$,
though $g_{rl}$ is always $O(\delta\log\frac{n}{\delta})$ and can be
$o(\delta\log\frac{n}{\delta})$ in some text families 
\citep{KNPtit23}. 
The MEM finding algorithm is adapted to find the MUMs between $P$ and $T$
within the same space and time complexities.

We also considered the extended problem of computing the $k$-MEMs of $P$, with
$k$ given at query time. The above complexities are generalized by multiplying 
the $\log^\epsilon n$ terms by $k$, while retaining the space. For larger $k$, 
we obtain $O(m^2\log^{2+\epsilon}n)$ time within
space $O(g)$ given a CFG of size $g$ that generates $T$, and $O(m\log m 
\log^{2+\epsilon}n)$ time within space $O(\delta\log\frac{n}{\delta})$.
We find $k$-rare MEMs, which generalize MUMs, within the same space and time
complexities of $k$-MEMs.

Our techniques are presented on a particular locally consistent grammar
\citep{KNOlatin22,KNOarxiv23} that yields the best complexities, but they would 
work on others too (possibly without the same worst-case time guarantees). 
We believe they could be successfully implemented on practical constructions of 
CFGs \citep{CNPjcss21} built with RePair \citep{LM00}
or of locally consistent grammars based on induced suffix sorting 
\citep{DNPspire21.2,NLGANjea21.2}. Further, even lacking theoretical 
guarantees, the algorithm for arbitrary RLCFGs will probably be competitive
if implemented on Lempel-Ziv based indices \citep{KN13,FKP18}, which are considerably
smaller than those based on grammars.

As we have shown in various applications, our development opens the door
to running, on parsing-based indices, algorithms that were reserved to
suffix-based ones. This is because the classic exact pattern matching based on
cutting points requires that the pattern to match is completely specified from
the beginning, therefore excluding problems where one must match parts of the
pattern as far as possible. The holy grail would be to simulate the suffix tree
functionality within $O(g)$ space, where $g$ is the size of the best grammar, 
from some relevant family, that represents the text, for example $g = 
O(\delta\log\frac{n}{\delta})$ with our locally-consistent grammars. Currently 
we can do this only in $O(\delta\log^2\frac{n}{\delta}\log\delta)$ space, by 
simulating the suffix tree in $O(r\log\frac{n}{r})$ space with the r-index 
\citep{GNPjacm19}, given that $r=O(\delta\log\frac{n}{\delta}\log\delta)$
\citep{KK20}. Note this is still a suffix-based index; we believe fundamentally
different techniques, truly based on grammar compression, are necessary to
remove the polylogarithmic redundancy in the space.

\section*{Acknowledgements} 

We thank Tatiana Starikovskaya, Tomasz
Kociumaka, and Adri\'an Goga for useful discussions. 

\section*{Declarations}

Funded by Basal Funds FB0001, ANID, Chile.

\bibliography{paper}

\end{document}